\documentclass[aps,pra,showpacs,preprint]{revtex4}
\usepackage{graphicx}
\usepackage{amsmath,amssymb,amsfonts}
\usepackage{dcolumn}
\usepackage[french,english]{babel}
\usepackage{epsfig}

\def\bb {\begin {equation}}
\def\ee {\end {equation}}

\def\f{\mbox{\boldmath $f$}}
\def\F{\mbox{\boldmath $F$}}
\def\a{\mbox{\boldmath $a$}}
\def\b{\mbox{\boldmath $b$}}
\def\c{\mbox{\boldmath $c$}}
\def\B{\mbox{\boldmath $\mathrm{B}$}}
\def\H{\mbox{\boldmath $\mathrm{H}$}}
\def\I{\mbox{\boldmath $\mathrm{I}$}}
\def\Om{\mbox{\boldmath $\Omega$}}
\def\R{\mbox{\boldmath $\mathrm{R}$}}
\def\S{\mbox{\boldmath $\mathrm{S}$}}

\def\Bpsi{\mbox{\boldmath $\Psi$}}
\def\BXi{\mbox{\boldmath $\Xi$}}
\def\BPhi{\mbox{\boldmath $\Phi$}}
\def\By{\mbox{\boldmath $\mathrm{Y}$}}
\def\Bd{\mbox{\boldmath $\mathrm{D}$}}

\begin{document}

\title{Time scaling with efficient time-propagation techniques for atoms and molecules in pulsed radiation fields }

\author{Aliou Hamido$^1$, Johannes  Eiglsperger$^{2,3}$, Javier  Madro\~{n}ero$^3$, Francisca 
              Mota-Furtado$^4$, Patrick O'Mahony$^4$, Ana Laura Frapiccini$^1$\footnote{On leave of absence from Divisi\'on   
              Colisiones At\'omicas, Centro At\'omico Bariloche, 8400 S.C. de Bariloche, Rio Negro, Argentina.} and Bernard 
              Piraux$^1$}

\affiliation{$^1$Institute of Condensed Matter and Nanosciences,  Universit\'e Catholique de  Louvain, \\
                            B\^{a}timent de Hemptinne, 2, chemin du cyclotron, B1348 Louvain-la-Neuve, Belgium.\\
                   $^2$Institut f\"ur Theoretische Physik, Universit\"at Regensburg, D-93040 Regensburg, 
                            Germany.\\
                   $^3$Physik Departement, Technische Universit\"at M\"{u}nchen, D-85747 Garching, 
                            Germany.\\
                   $^4$Department of Mathematics, Royal Holloway, University of London, Egham,\\
                             TW20 0EX Surrey, United Kingdom.}
\date{\today}

\begin{abstract}
We present an {\it ab initio} approach to solve the time-dependent Schr\"odinger equation to treat electron and photon impact multiple ionization of atoms or molecules. It combines the already known time scaled coordinate method with a new high order time propagator based on a predictor-corrector scheme. In order to exploit in an optimal way the main advantage of the time scaled coordinate method namely that the scaled wave packet stays confined and evolves smoothly towards a stationary state the modulus square of which being directly proportional to the electron energy spectra in each ionization channel, we show that the scaled bound states should be subtracted from the total scaled wave packet. In addition, our detailed investigations suggest that multi-resolution techniques like for instance, wavelets are the most appropriate ones to represent spatially the scaled wave packet. The approach is illustrated in the case of the interaction of an one-dimensional model atom as well as atomic hydrogen with a strong oscillating field.

\end{abstract}

\maketitle

\section{Introduction}

During the last few years, substantial progress regarding the development of new XUV sources, has been made in two directions. On the one hand, high order harmonic generation has been used to produce  attose\-cond pulses of which the duration is of the order of the characteristic time scale of the inner-shell electron dynamics in atoms and molecules \cite{Paul01}. On the other hand, free electron lasers \cite{And00}  are now operating at unprecedentedly high peak intensities in the far-X-ray regime. These developments have opened the route to the exploration of non linear processes in the short-wavelength limit. At present,  processes such as multiphoton multiple ionization  of atoms and molecules are the focus of many experimental and theoretical studies with a view to understanding the subtle role of the electronic correlations.\\

Within this context, there is clearly a need for reliable theoretical and numerical methods that provide accurate solutions of the time-dependent Schr\"odinger equation (TDSE). To this end however, it is necessary to overcome the following four main difficulties. (i) The continuum components of the wave packet expand in a rapidly increasing volume of space, thereby requiring very extended spatial grids or basis functions in order to avoid artificial reflections from the numerical boundaries. (ii) Increasingly large spatial phase gradients develop within the wave packet with time, demanding very dense grids or large basis sizes. (iii)  Solving the TDSE on a spatial grid or in a basis of square integrable functions leads to a stiff system of equations which, in principle, makes explicit time propagators unstable. Finally, (iv) the direct extraction of the information on the multi-electron continua from the wave packet necessitates the knowledge of the asymptotic behavior of the corresponding wave function.\\

The existing time dependent approaches have been mainly used to study single and double ionization of two-electron atoms and molecules by intense ultrashort radiation fields. In the low frequency regime where the calculations are extremely challenging,  Smyth {\it et al.} \cite{Smyth98} have developed a fully numerical method to solve the TDSE.  It has provided valuable qualitative information on the role of the electronic correlations and the so-called rescattering process \cite{Parker06}. In the high frequency regime where one or two photons are involved in the ionization process, there are presently two types of treatment  to solve the TDSE: the treatments  based on standard methods of collision theory and the close-coupling approaches. In the former case, the wave packet is time propagated on an extended spatial grid during a period of time that is much larger than the pulse duration. The Fourier transform of the wave packet provides a scattered wave function which is then analyzed by means of time independent methods. Palacios {\it et al.} \cite{Pala09} use the Exterior Complex Scaling  (ECS) technique which maps an outgoing wave into a vanishing wave outside a physically unaltered region allowing  the extraction of the relevant information on the various ionization processes without the necessity of knowing the asymptotic behavior of the wave function associated to the multiple continua. Recently, Malegat {\it et al.} \cite{Maleg10} applied the Hyperspherical R-Matrix with Semiclassical Outgoing Waves (HRM-SOW) method to calculate the various ionization yields. In this method, the scattered wave is propagated semiclassically with res\-pect to the hyperradius, all the way to the asymptotic region where the various ionization channels are decoupled.\\

\newpage
Many approaches based on a close-coupling method have been developed. They essentially differ by the way the information on the ionization processes is extracted from the wave packet. The most common way is to pro\-pagate the wave packet freely after the interaction with the pulse, until it reaches a spatial region where the ionization channels are assumed to be decoupled. It is then projected onto an uncorrelated product of Coulomb functions in each of the ionization channels \cite{Colgan01,Feist08,Guan08,Guan09}.  Instead of using uncorrelated products of Coulomb functions, Ivanov and Kheifets  \cite{Ivanov08} project the wave packet on continuum state wave functions obtained by means of the Convergent Close
Coupling (CCC) method which takes into account electron correlations in an approximate way. A different procedure has been deve\-loped by Foumouo {\it et al.} \cite{Foum06}. Since the asymptotic behavior of the single continuum wave function is known, it is convenient to calculate the total probability for double ionization by subtracting the total probability for single ionization from the all-inclusive probability for breakup which in turn can be calculated without any reference to the boundary conditions. In order to calculate the total and partial probabilities for single ionization, they use the Jacobi-matrix method to generate a multichannel scattering wave function that describes the single continuum. The projection of the wave packet on this scattering wave function is performed just at the end of the interaction of the two-electron system with the radiation pulse.\\

Finally, Lysaght {\it et al.} \cite{Lysag08,Lysag09} have recently ini\-tiated the development of a Time-Dependent $R$-Matrix (TDRM) approach to describe complex multielectron atoms and atomic ions in intense ultrashort radiation pulses. This approach consists in time propagating the atomic wave function in the presence of the radiation field both in the internal and external $R$-matrix regions.\\

The present approach combines the Time Scaled Coordinate (TSC) method with a high order fully implicit predictor-corrector scheme for the time propagation. The time dependent scaling of the radial electronic coordinates together with a phase transformation of the wave packet allow for  ``freezing" the spatial expansion of the wave packet in the new representation while removing fast oscillations due to the increasingly large spatial phase gradients that develop with time. This method is in fact equivalent to using a time-dependent basis that expands in the same way as the wave packet itself. This idea of time scaling the coordinates is not new and has been widely exploited in many different fields of physics. In 1979,  Burgan {\it et al.} \cite{Burgan79} studied the Schr\"odinger equation for a multidimensional quantum harmonic oscillator with time-dependent frequencies. By introducing an appropriate time-dependent scaling of the spatial coordinates, they were able to transform the problem to a free particle motion and to derive an exact analytical solution. Later on, Manfredi {\it et al.} \cite{Manfred93,Mola93} introduced a time-dependent scaling of both space and time variables to  ``freeze" the expansion into a vacuum of both a one-dimensional, collisionless, two-species classical plasma and a quantum electron gas in planar geometry. In atomic and mole\-cular physics, Solov'ev {\it et al.}  \cite{Solo85} and later on,  Ovchinnikov {\it et al.} \cite{Ovchi04}  treated the Coulomb three-body problem, in particular ion-atom and atom-atom collisions,  within a proper adiabatic representation by time scaling the internuclear distance. More recently, the TSC method has been used by Sidky {\it et al.} \cite{Sidky00} and Derbov {\it et al.} \cite{Derbov03} to treat the interaction of a model atom and molecule with an electromagnetic pulse and by Serov {\it et al.} \cite{Serov01,Serov07} to study electron impact single and double ionization of helium and more recently, double photoionization of two-electron atomic systems \cite{Serov08}. The TSC method is somehow an extension of a self-similarity analysis which has been introduced recently in astrophysics \cite{Falize10}. By an appropriate scaling of all variables entering the equations governing the dynamics of a very large hydrodynamic system, the rescaled equations are identical to the original ones. This allows to define dual equivalent systems, the first one characterized by very long time and parsec length scales and the second one, characterized by very short time and small length  scales allowing its study at the laboratory scale. Finally, the TSC method has been used to study the expansion of a Bose-Einstein condensate following the switch off of the trap \cite{Castin96, Pita04}.\\

In the case of the interaction of an atom or a molecule with an electromagnetic pulse, the TSC method 	effectively confines the expansion of the scaled wave packet within a finite space of controllable size so that the evolution of this scaled wave packet can be followed over very long periods of time . Furthermore, it has been shown \cite{Serov01,Derbov03,Roud05} that a long time after the end of the interaction of the atom or the molecule with the pulse, the energy spectrum of the ejected electrons is simply proportional to the modulus square of this scaled wave packet. The confinement of the scaled wave packet is due to three factors: the presence of an harmonic potential, the narrowing of the atomic potential and the increase of the effective mass of the electrons with time. This means that the effective de Broglie wavelength of these electrons decreases. In other words,  the TSC method introduces different length scales in the problem. This has two important consequences. First, an optimal spatial description of the scaled wave packet requires multi-resolution techniques and second, it increases significantly the stiffness of the system of first order differential equations to solve for the time propagation of the wave packet. By this, it is meant that the time step rapidly decreases with increasing size of the system
\cite{Lambert91}. In this contribution, we describe a seventh order fully implicit predictor-corrector scheme. The predictor is the fifth order explicit method of Fatunla \cite{Fatun78,Fatun80} while the corrector is a seventh order fully implicit  Radau method \cite{Butcher64}. In principle, an implicit scheme requires solving large systems of algebraic equations at each time step. However,  the accuracy of Fatunla's method is high enough \cite{Madro09,Madro10} to allow the use of an iterative procedure, the biconjugate gradient algorithm, to solve the large systems of algebraic equations at the corrector level. In other words, only matrix-vector products are needed, allowing a deep parallelization of the computer code.\\

This contribution is organized as follows. In the first section after this introduction, we describe the TSC method in detail. For the sake of illustration, we consider the interaction of an one-dimensional system modeled by a Gaussian potential and interacting with a cosine square electromagnetic pulse. First, we examine the different reasons for the confinement of the scaled wave packet. Then, we study various spatial representations of this scaled wave packet and study its behaviour at various times after the pulse has ceased to interact with the model atom. In the next section, we describe our time propagation method. We first start with the explicit Fatunla's method and then discuss in detail the predictor-corrector scheme. The third section is devoted to the calculation of the energy spectrum. We derive an analytical expression in the case of our model atom and for atomic hydrogen. Results for both cases are presented and discussed in detail. The last section is devoted to conclusions and perspectives. Unless stated otherwise, atomic units are used throughout this paper.

\section{The time scaled coordinate method}
\subsection{Outline of the method}
Our one-dimensional model that serves as an illustration for describing the TSC method consists of an electron initially bound in a Gaussian potential and interacting with a cosine square electromagnetic pulse. The TDSE that governs the dynamics of the electron is:
\bb
\mathrm{i}\frac{\partial}{\partial t}\Psi(x,t)=\left(H_0(x)+H_I(x,t)\right)\Psi(x,t).
\ee
The atomic Hamiltonian $H_0(x)$ is given by:
\bb
H_0(x,t)=-\frac{1}{2}\frac{\partial^2}{\partial x^2}+V(x), 
\ee
where:
\bb
V(x)=-V_0\mathrm{e}^{-\beta x^2}.
\ee
By adjusting the parameters $V_0$ and $\beta$ that fix the depth and the width of the Gaussian potential, we can easily vary the number of bound states. In all the calculations we perform, we always assume that the model atom is initially in its ground state. Within the dipole approximation and in the velocity form, the interaction Hamiltonian $H_I(x,t)$ writes:
\bb
H_I(x,t)=-\mathrm{i}A_0f(t)\sin(\omega t+\varphi)\frac{\partial}{\partial x}.
\ee
$A_0$ is the amplitude of the vector potential that is polarized along the $x$-axis. $\omega$ is the frequency and $\varphi$ the carrier phase. $f(t)$ is the pulse envelope defined as follows:
\bb
f(t)=
\left\{
\begin{tabular}{p{2.5cm}l}
$\cos^2(\frac{\pi}{\tau}t)$, & $|t|\le \frac{\tau}{2}$\\ \\
0, & $|t|> \frac{\tau}{2}$
\end{tabular}
\right .
\ee
The total pulse duration $\tau=2\pi n_{\mathrm{c}}/\omega$ where $n_{\mathrm{c}}$ is an integer giving the number of optical cycles. The fact that $n_{\mathrm{c}}$ is an integer is important since it ensures that the electric field has no static components.\\

According to the TSC method \cite{Sidky00}, we introduce the scaled coordinate $\xi$ given by:
\bb
\xi=\frac{x}{R(\sigma)},
\ee
where $R(\sigma)$ is an arbitrary scaling function. It is important to stress that $\sigma$ is just a parameter. In the following, we assume that it coincides with the time $t$. We write the scaled wave packet as follows:
\bb
\Phi(\xi,t)=\sqrt{R}\;\mathrm{e}^{-\mathrm{i}R\dot R\xi^2}\Psi(R\xi,t),
\ee
where the dot indicates the time derivative. The factor $\sqrt{R}$ ensures that this scaled wave packet is correctly normalized. The phase transformation absorbs the fast oscillations of the unscaled wave packet during its time evolution. The scaled wave packet $\Phi(\xi,t)$ satisfies the following TDSE:
\bb
\mathrm{i}\frac{\partial}{\partial t}\Phi(\xi,t)=\left[-\frac{1}{2}\frac{\partial^2}{R^2\partial\xi^2}+V(R\xi)-\mathrm{i}\frac{A_0}{R}f(t)\sin(\omega t+\varphi)\frac{\partial}{\partial\xi}+\frac{1}{2}R\ddot R\xi^2\right]\Phi(\xi,t).
\ee
Since the idea behind the TSC method is to build a time-dependent basis that expands in the same way as the wave packet, it is expected that if this expansion is accelerated, non inertial forces should appear. This explains the presence of the harmonic potential in the above TDSE. When this expansion occurs at constant velocity, {\it i.e.} when the scaling function is linear with time, the harmonic potential disappears. For $\ddot R>0$, this potential confines the wave packet in a finite space. In fact, in the absence of external fields, the spectrum of the operator in square brackets in Eq. (8), becomes  purely discrete \cite{Serov07}. An analysis of Eq. (8) shows that the scaling transformation (6) introduces an electron effective mass that is proportional to $R^2$. Finally, we see that when $R$ increases, the Gaussian potential narrows with time. This shrinking mainly affects the bound state components of the wave packet. Note that in the case of a Coulomb potential it is the effective electric charge that goes to zero with increasing values of $R$.\\

Before analysing in more details the different factors that lead to the confinement of the scaled wave packet, let us examine the scaling function $R(t)$. As stressed in \cite{Sidky00}, this function 
\begin{figure}[h]
\begin{center}
\includegraphics[width= 9cm,height = 7cm]{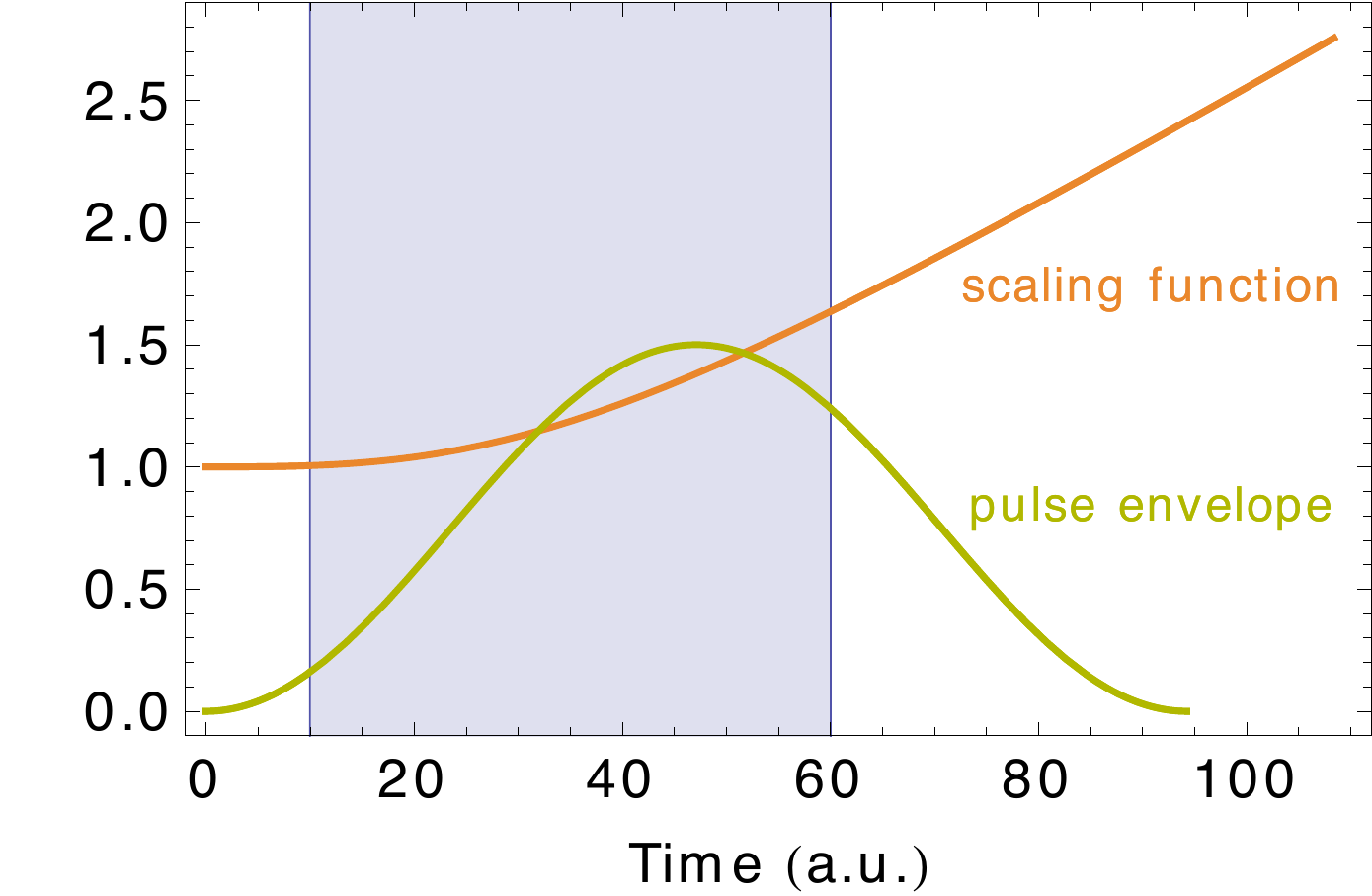}
\caption{(Color online) Scaling function and pulse envelope as a function of time. For the scaling function, $n=3$, $t_{\mathrm{sc}}=t_{\mathrm{initial}}$ and $R_{\infty}=0.025$. The pulse has a sine square envelope and a total duration of 94 a.u..}
\end{center}
\end{figure}
is arbitrary and chosen to facilitate the numerics. It must however satisfy 
a few constraints. It should be real, larger than one and equal to one from $t=t_{\mathrm{initial}}$ corresponding to the beginning of the interaction until $t=t_{\mathrm{sc}}$ where the scaling starts. In addition, for large time $t$, $R$ should tend to $R_{\infty}t$ where $R_{\infty}$ is what we call the asymptotic velocity. This ensures that for large times, the scaled wave packet becomes stationary. In the present calculations, we define $R(t)$ as follows:
\bb
R(t)=
\left\{
\begin{tabular}{p{5cm}l}
1, & $t\le t_{\mathrm{sc}}$\\ \\
$\{1+[R_{\infty}(t-t_{\mathrm{sc}})]^n\}^{\frac{1}{n}}$, & $t>t_{\mathrm{sc}}$.
\end{tabular}
\right .
\ee
This form with $n=4$ has been used by several authors \cite{Sidky00,Derbov03,Roud05}. It leads to a smooth transition
to the linear regime where the harmonic potential is switched off. The scaling function for $n=3$, $t_{\mathrm{sc}}=t_{\mathrm{initial}}$ and $R_{\infty}=0.025$ together with a sine square pulse envelope are shown as a function of time in Fig. 1. In the shaded region, between 10 and 60 a.u. of time, $\ddot R(t)$ is significantly larger than zero. In that case, the confinement of the scaled wave packet results predominantly from the presence of the harmonic potential. For times $t>60$, the harmonic potential is smoothly switched off but the confinement of the scaled wave packet subsists because of the electron effective mass which rapidly increases. As soon as the scaling starts (at $t=t_{\mathrm{sc}}$), the atomic bound state wave functions start to 
\begin{figure}[h]
\includegraphics[scale=0.23]{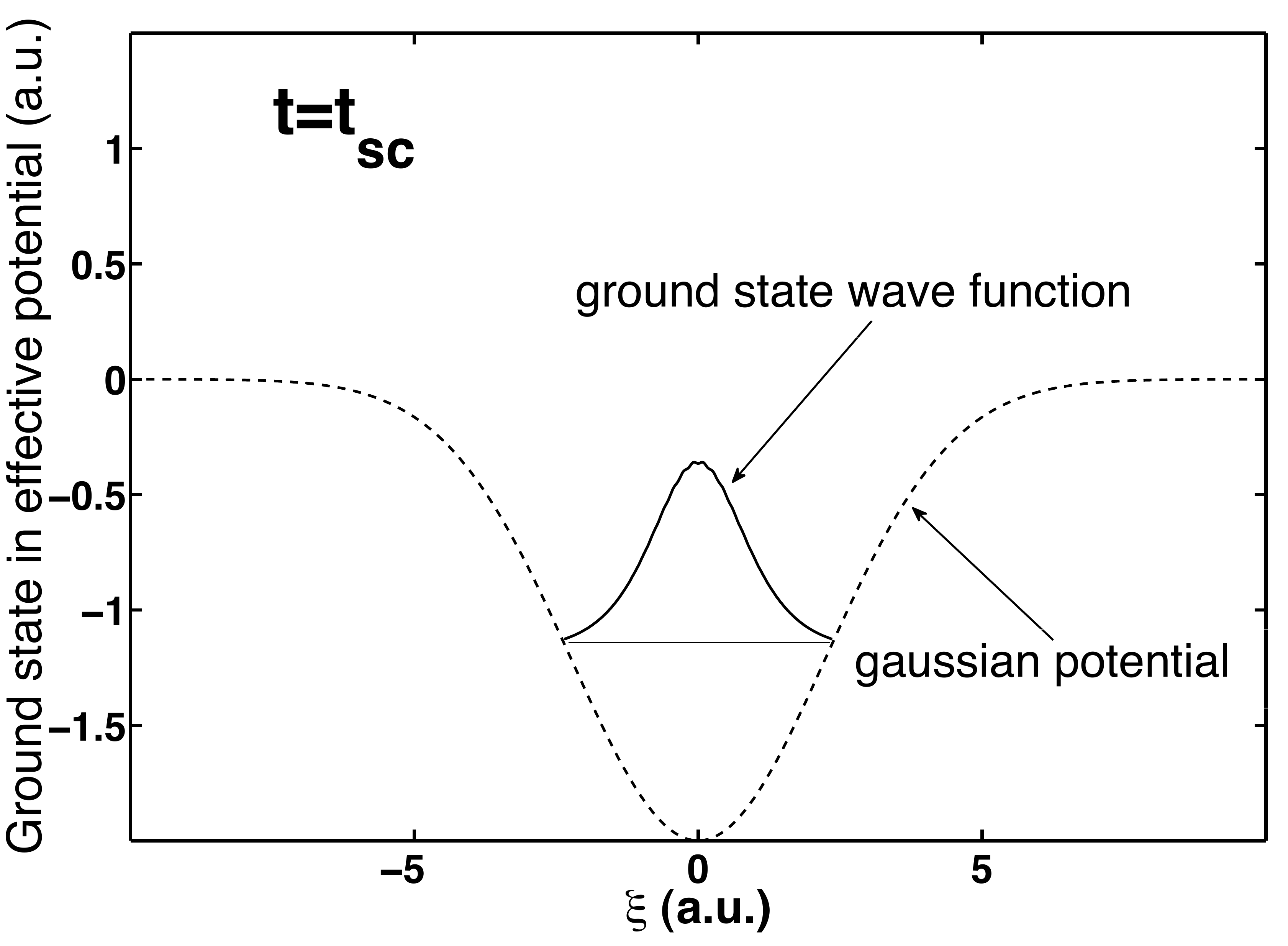} \includegraphics[scale=0.23]{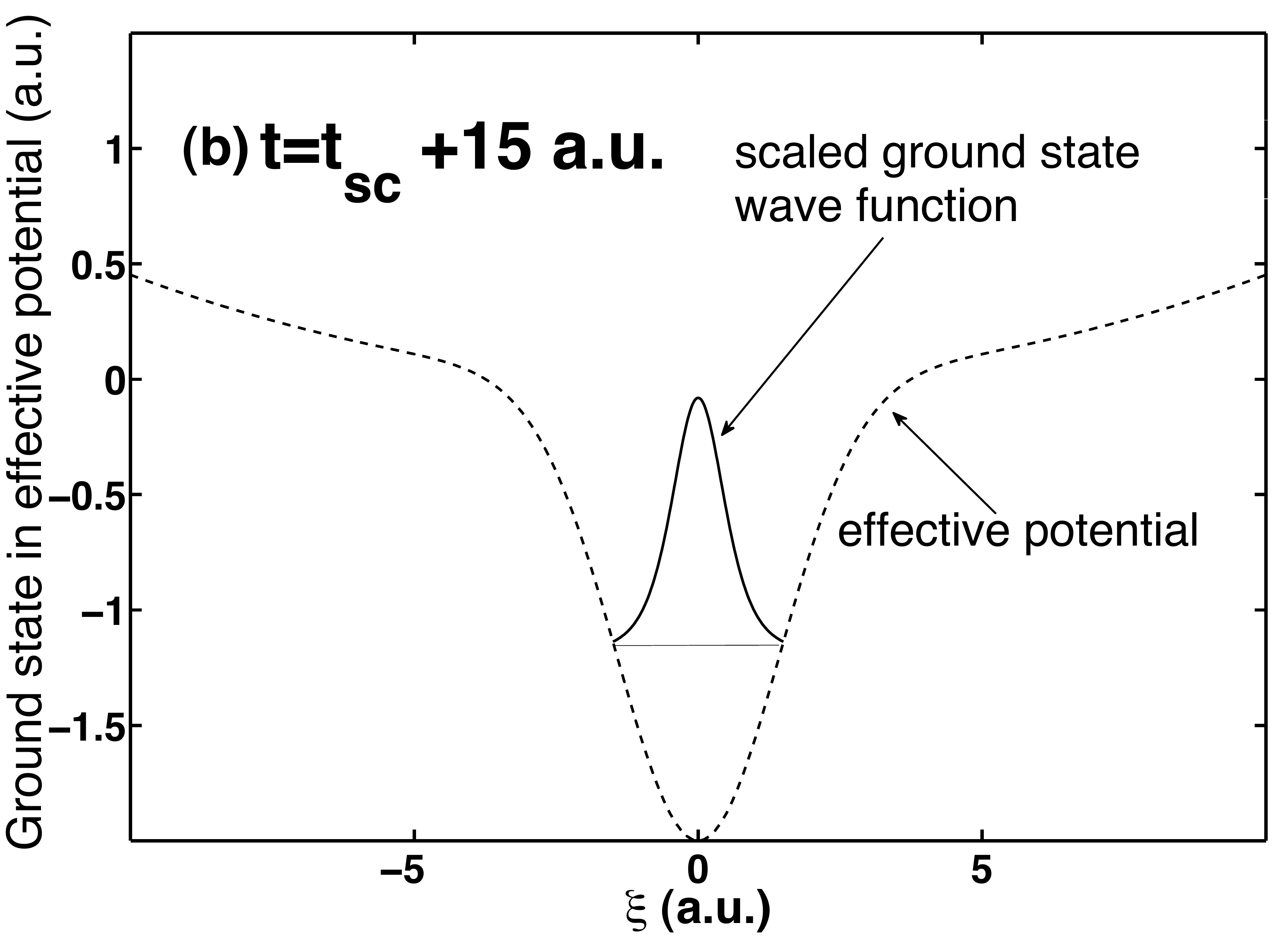}\includegraphics[scale=0.23]{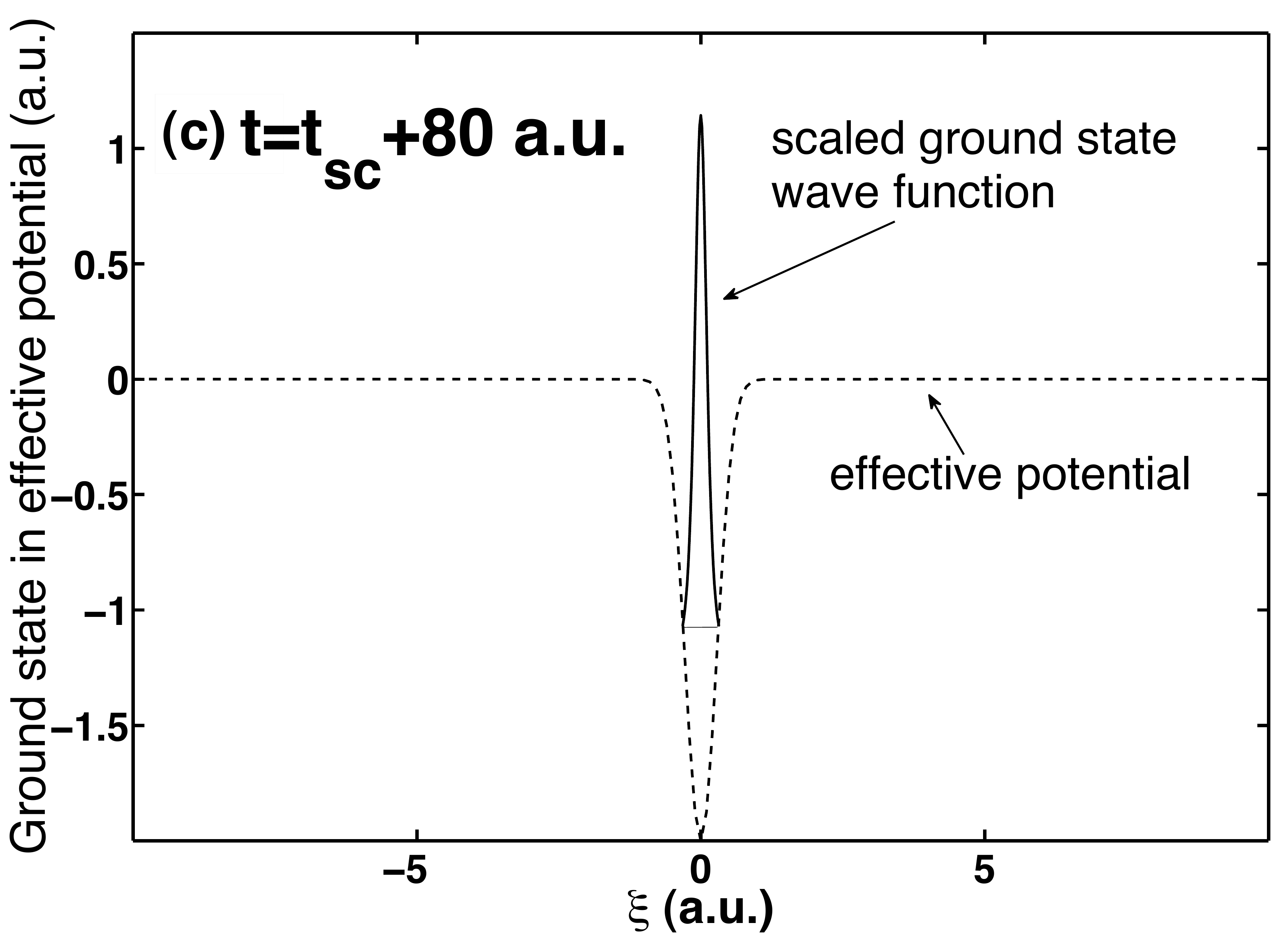}
\caption{Effective potential  namely the sum of the scaled Gaussian potential and the harmonic one together with the scaled ground state wave function versus the scaled variable $\xi$. Three times are considered: (a) $t=t_{\mathrm{sc}}$ corresponding to the time where scaling is switched on (b) $t=t_{\mathrm{sc}}+15$ a.u. and $t=t_{\mathrm{sc}}+80$ a.u..
The asymptotic velocity is equal to 0.1 a.u. and $n=4$ (see Eq. (9)).}
\end{figure}
shrink because of the narrowing of the atomic potential. In the case of the Gaussian potential it is easy to show that the width is inversely proportional to $R$. This is illustrated in Fig. 2 where we show the scaled ground state wave 
function as well as the effective potential which is the sum of the scaled Gaussian potential and the harmonic one as a function of the scaled variable $\xi$, for three different values of $t$: $t_{\mathrm{sc}}$, $t_{\mathrm{sc}} +15$ a.u. and $t_{\mathrm{sc}}+80 $ a.u.. Note that at $t=t_{\mathrm{sc}}$, the effective potential reduces to the Gaussian potential. The fact that the scaled ground state wave function shrinks rapidly is due to the relatively high value of the asymptotic velocity which, in the present case, is equal to 0.1 a.u.. As $n=4$ (see Eq. (9)), $t=t_{\mathrm{sc}}+80$ a.u. corresponds to a time for which the scaling function has reached its linear regime. It is important to mention that the results given in Fig. 2 require only a partial diagonalization of the atomic Hamiltonian and no time propagation. Therefore, they provide an easy way to control the fineness of the spatial discretization necessary to maintain the accuracy of the time propagation.

\subsection{Spatial representation of the scaled wave packet}
The optimal way of describing the wave packet in space is based on a multi-resolution analysis \cite{Daub92}. The general idea is to define different resolution levels in various regions of space through the introduction of several grids with a density of mesh points that increases from one grid to the next one in the spatial regions of interest. These techniques will be analysed in detail in a forthcoming publication. Here, we use two different spectral methods. The first one consists in developing the wave packet on $\mathcal{L}^2$ integrable functions namely Hermite-Sturmian functions in the case of our one-dimensional model and Coulomb-Sturmian functions in the case of atomic Hydrogen. The second method uses a basis of B-splines built on a non-uniform grid with an exponential sequence of breakpoints. These two methods that are far from being optimal, have the merit to be easily implemented and to clearly show that  one cannot dissociate the spatial discretization problem from the time propagation.\\

Let us now briefly  describe our first spectral method. In the case of our one-dimensional model, we expand the total wave packet in a finite basis of Hermite-Sturmian functions as follows:
\bb
\Psi(x,t)=\sum_{n=0}^Na_n(t)\varphi_n^{\alpha}(x),
\ee
where $a_n(t)$ is the expansion coefficient and $\varphi_n^{\alpha}(x)$, the Hermite-Sturmian function given by:
\bb
\varphi_n^{\alpha}(x)=\left(2^nn!\sqrt{\frac{\pi}{\alpha}}\right)^{-\frac{1}{2}}\mathrm{e}^{-\frac{\alpha}{2}x^2}H_n(\sqrt{\alpha}x).
\ee
The elements of the matrices associated to all operators present in the scaled and unscaled Hamiltonians can be calculated analytically except for the Gaussian potential. In this latter case, a Gauss-Hermite quadrature provides exact results for a sufficient number of abscissae. $\alpha$ is a dilation parameter that determines the resolution of the basis. Indeed, a large value of $\alpha$ gives a good description of a wave packet which exhibits sharp variations close to the origin. In that case however, a large value of $N$, the number of basis functions, is necessary if the extent of this wave packet is significant. By contrast, a small value of $\alpha$ allows a good description of the wave packet over much larger distances but $N$ has to be extremely large again if sharp variations of the wave packet occur. \\

\newpage
In the case of atomic hydrogen where we use spherical coordinates, we write the wave packet as follows:
\bb
\Psi(\vec{r},t)=\sum_{n,l,m}a_{n,l}(t)\frac{S^{\kappa}_{n,l}(r)}{r}Y_{l,m}(\theta,\phi),
\ee
where $Y_{l,m}(\theta,\phi)$ is a spherical harmonic and $S^{\kappa}_{n,l}(r)$ the Coulomb-Sturmian function given by:
\bb
S^{\kappa}_{n,l}(r)=N^{\kappa}_{n,l}\;r^{l+1}\mathrm{e}^{-\kappa r}L^{2l+1}_{n-l-1}(2\kappa r).
\ee
$N_{n,l}^{\kappa}$ is a normalization factor (see \cite{Foum06} for details) and $L_{n-l-1}^{2l+1}(2\kappa r)$ a Laguerre polynomial. The index $n$ varies from $l+1$ until $\infty$. The Coulomb Sturmian functions form a complete and discrete set of $\mathcal{L}^2$ integrable functions that are the exact solutions of the stationary Schr\"odinger equation for a single electron in the Coulomb field of a nucleus of charge $Z$ for selected values of $\kappa$. As a result, these functions are well adapted to describe the energy spectrum of atomic hydrogen as well as its behavior in presence of an external field. In fact, the matrices associated to the corresponding Hamiltonian are banded with a narrow bandwidth (three diagonals). As in the case of the Hermite Sturmian functions, a given basis of Coulomb Sturmian functions is characterized by a fixed value of the dilation parameter $\kappa$. An interesting idea in the spirit of the multi-resolution approaches is to consider a set of different values of $\kappa$ within a given basis in order to take into account the various length scales in the problem. Despite the fact that the introduction of different values of $\kappa$ makes the basis numerically overcomplete thereby requiring the elimination of the linearly dependent eigenvectors of the overlap matrix, this idea turned out to be extremely successful to generate the singly and doubly excited states of helium \cite{Foum06,Eigl09,Eigl10a,Eigl10b,Eigl10c}. Note that in the case of atomic hydrogen, the time dependent scaling of the radial coordinate is equivalent to introducing a time dependent $\kappa$.\\

In our second method to treat our one-dimensional model, we expand the wave packet in a basis of B-splines \cite{Bach01}
\bb
\Psi(x,t)=\sum_{i=1}^Nc_i(t)B_i^k(x),
\ee
where $B_i^k(x)$ is a B-spline of order k. In the present calculations, $k=7$. In order to calculate the B-splines, we use an exponential sequence of breakpoints. In practice, we proceed as follows. We adjust the asymptotic velocity and the time $t_{\mathrm{sc}}$ at which the scaling starts so that the scaled wave packet is confined in a relatively small interval [-$x_{\mathrm{max}}\;,+x_{\mathrm{max}}$]. We then define two symmetrical exponential sequences of breakpoints in the intervals [-$x_{\mathrm{max}}\;,0$] and [0\;,$\;+x_{\mathrm{max}}$]. In the  interval [0\;,$\;+x_{\mathrm{max}}$] for instance, the breakpoints $\xi_i$ are given by:
\bb
\xi_i=x_{\mathrm{max}}\left(\frac{\mathrm{e}^{\gamma(\frac{i-1}{N-1})}-1}{\mathrm{e}^{\gamma}-1}\right)\;\;\;\;\;\;\;\;\;\;\;\;\;\;\;\;\;\;\;\;\;i=1,...,N.
\ee
Typically, we have $x_{\mathrm{max}}=35$, $\gamma=5$ and $N$ of the order of 100. In this case, breakpoints accumulate symmetrically around 0 in order to describe properly the shrinking of the bound states. However, this accumulation of breakpoints introduces high frequencies in the problem. In this B-spline basis, the energy spectrum of the unscaled atomic Hamiltonian contains very high positive eigenenergies of the order of 3000 a.u.. If the wave packet is time propagated in the B-spline basis, each component of the B-splines contains these high frequencies making the problem extremely stiff and leading to a dramatic decrease of the time step. Note that  this difficulty can be avoided by propagating the scaled wave packet in the atomic basis {\it i.e.} the basis in which the unscaled atomic Hamiltonian is diagonal. In that case however, it is necessary to solve a generalized eigenvalue problem that could be time consuming in the case of a more complex system like He or H$_2$.\\

\subsection{Evolution of the scaled wave packet}
Before examining the time evolution of  scaled  wave packets, it is instructive to analyze the behavior of the unscaled ones. We first consider the case of a Gaussian potential with $V_0=1$ a.u. and $\beta=1$ a.u.. This potential has only one bound state the energy of which is equal to -0.477 a.u.. This model atom interacts with a cosine square laser pulse of peak intensity $I_{\mathrm{peak}}=10^{13}$ W/cm$^2$ and frequency $\omega=0.7$ a.u.. The total duration of the pulse is 6 optical cycles. In Fig. 3, we show the real part of the wave packet at time $t=t_{\mathrm{final}}$ {\it i.e.} at the end of the pulse (dark grey curve) and 300 a.u. of time later (light grey curve). We actually represent the ionized wave packet {\it i.e.} the total wave packet without its bound state component. 
\begin{figure}[h]
\begin{center}
\includegraphics[width=12cm,height=10cm]{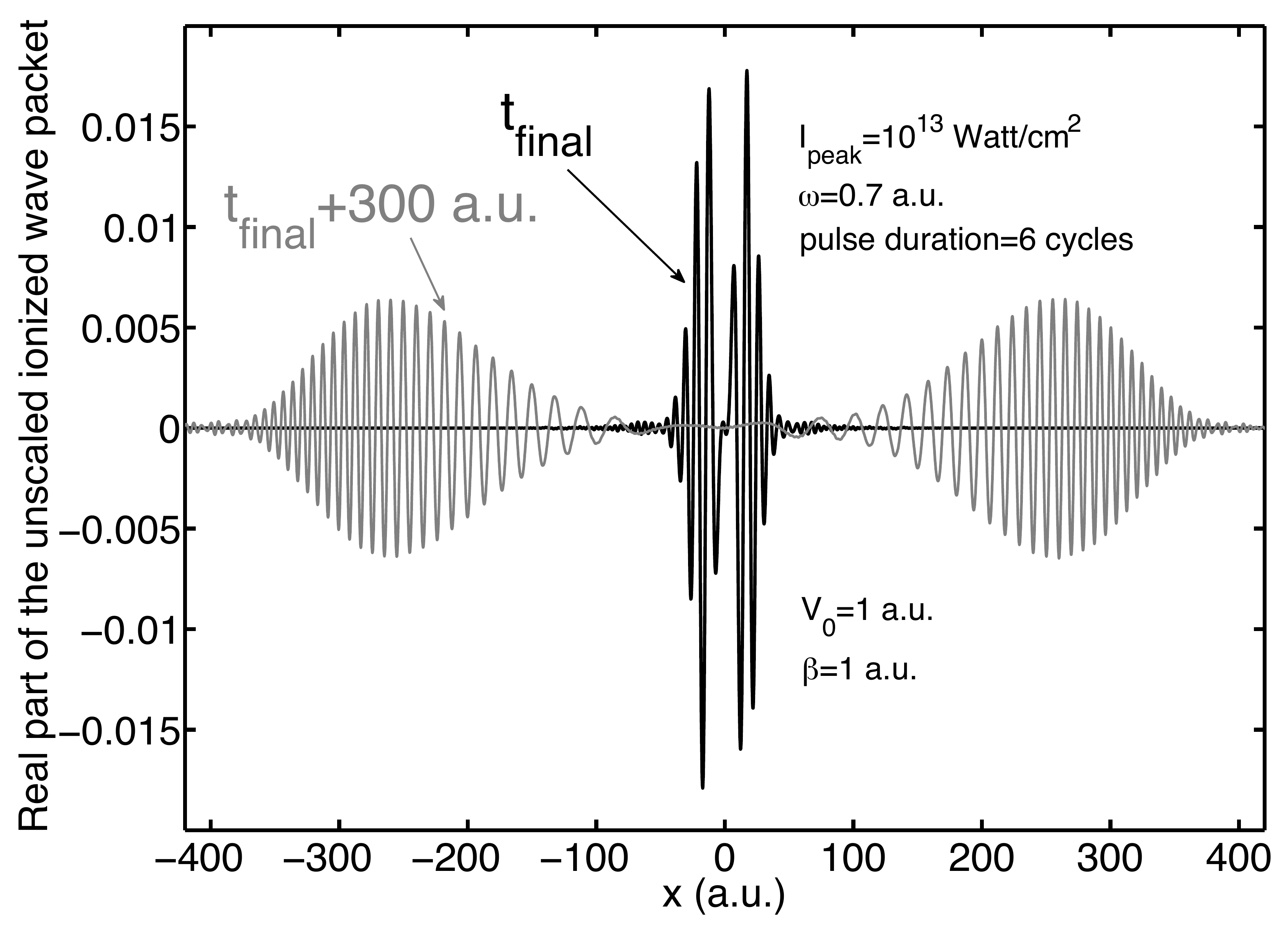} 
\caption{Real part of the unscaled ionized wave packet resulting from the interaction of our one-dimensional model atom with a cosine square pulse of $10^{13}$ Watt/cm$^2$ peak intensity and 0.7 a.u. photon energy. The total duration of the pulse is 6 optical cycles. The Gaussian potential depth $V_0=1$ a.u. and $\beta=1$ a.u.. The real part of the ionized wave packet is shown at time $t=t_{\mathrm{final}}$ {\it i.e.} at the end of the pulse (dark grey curve) and at time $t=t_{\mathrm{final}}+300$ a.u. (light grey curve)  . }
\end{center}
\end{figure}
We clearly see that at time $t=t_{\mathrm{final}}+300$ a.u., the real part of the ionized wave 
packet exhibits a much larger number of oscillations than at $t=t_{\mathrm{final}}$, 
the end of the pulse. In particular, we clearly see the presence of a chirp for $t=t_{\mathrm{final}}+300$. As stressed by Sidky {\it et al.} \cite{Sidky00}, this results from the phase gradients that rapidly develop 
\begin{figure}[h]
\begin{center}
\includegraphics[scale=0.29]{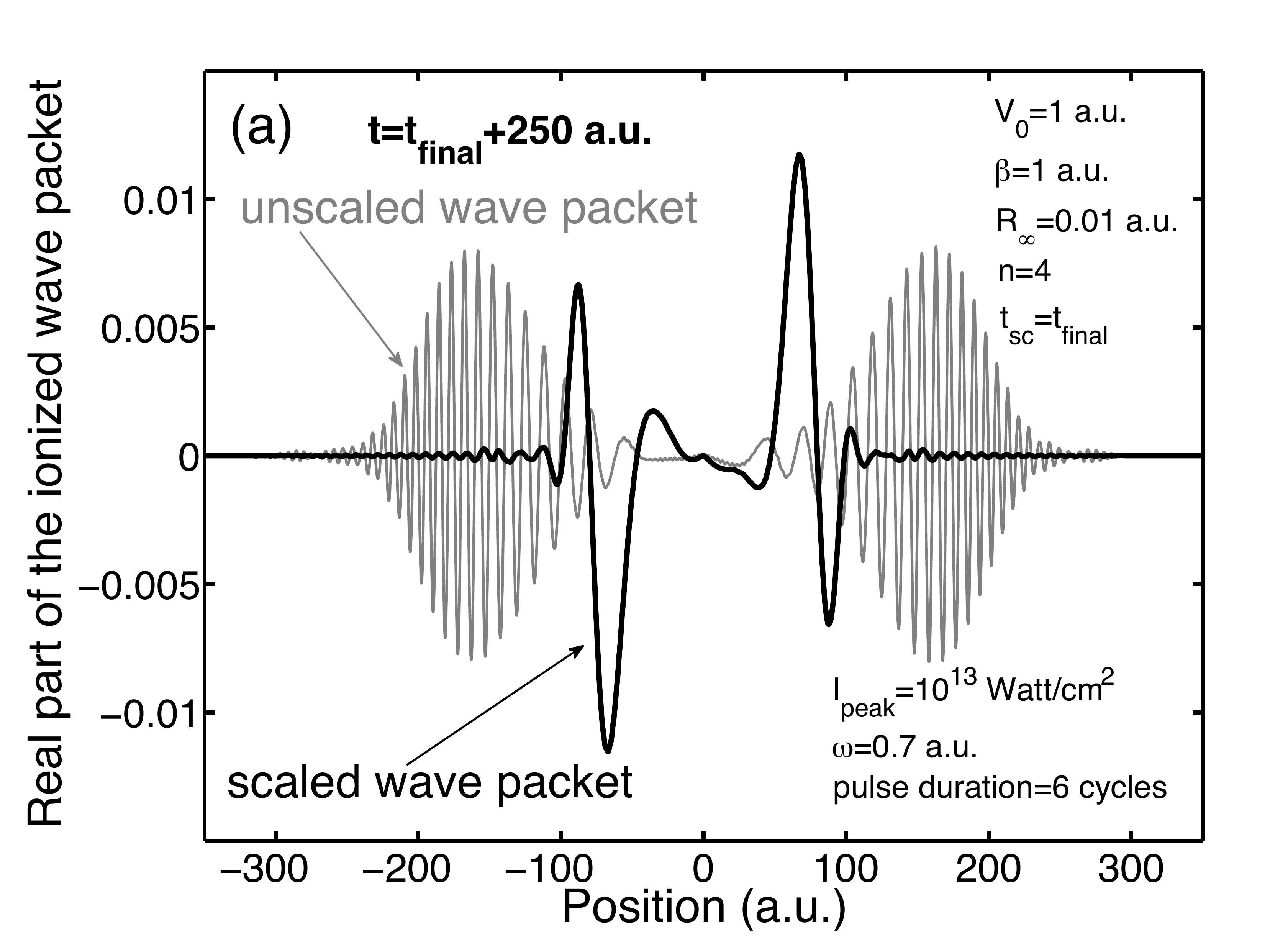} \includegraphics[scale=0.28]{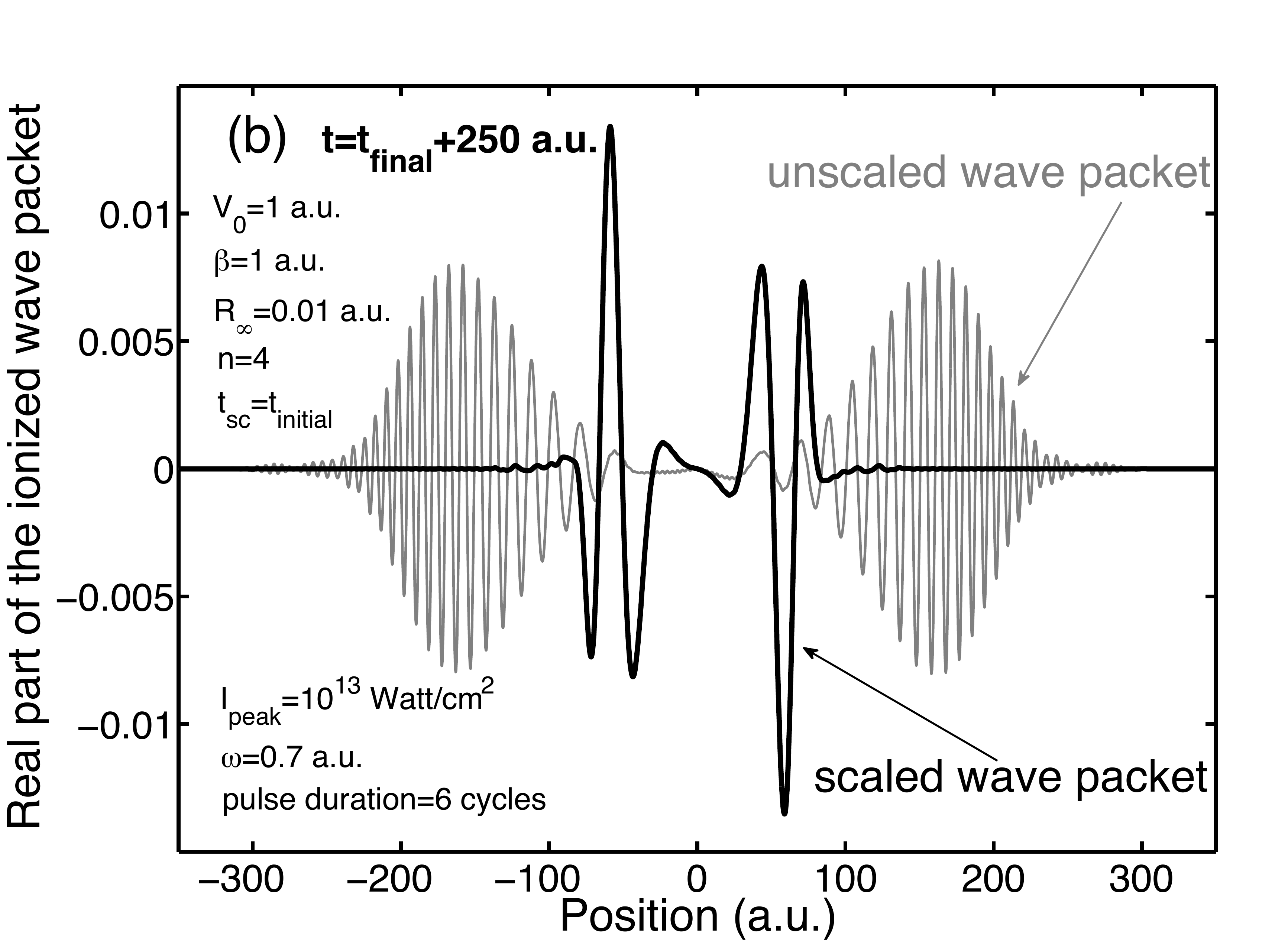}
\caption{Real part of both the scaled (dark grey curve) and unscaled (light grey curve) ionized wave packets resulting from the interaction of our one-dimensional model atom with the same pulse as in Fig. 3. The real part of these wave packets is calculated at time $t=t_{\mathrm{final}}+250$ a.u. and is represented as a function of the position: $x$ in the case of the unscaled wave packet and $\xi$ in the case of the scaled wave packet. The parameters of the Gaussian potential are the same as in Fig. 3. In the case of the scaled wave packet, the asymptotic velocity $R_{\infty} = 0.01$ a.u., the parameter $n$ of the scaling function (see Eq. 9) is equal to 4 and the scaling is switched on (a) at the end of the interaction and (b) at the beginning of the interaction of our model atom with the pulse.}
\end{center}
\end{figure}
over large distances or in other words, from the fact that the front edge of the wave packet is moving faster than the inner part. Let $\Psi(x,t_{\mathrm{final}})$ be the wave packet created at the end of the pulse. At any later time $t$, we have:
\bb
\Psi(x,t)=\int_{-\infty}^{\infty}\mathrm{d}x'\;\mathcal{G}(x,x',t)\Psi(x',t_{\mathrm{final}}),
\ee
where the Green function $\mathcal{G}(x,x',t)$ for a free electron is given by:
\bb
\mathcal{G}(x,x',t)=\sqrt{\frac{1}{2\mathrm{i}\pi t}}\mathrm{e}^{\mathrm{i}(x-x')^2/2t}.
\ee
This means that the phase increases quadratically with the distance. In the present calculations, we need to use a basis of 1000 Hermite Sturmian functions of parameter $\alpha=0.01$ to accurately reproduce all the oscillations over about 300 a.u. around the origin.
In Fig.4a, we compare for $t=t_{\mathrm{final}}+250$ a.u., the real part of both the scaled and unscaled ionized wave packets for the same pulse and Gaussian potential parameters as in Fig. 3. In this particular case, the scaled ionized wave packet is obtained as follows. The unscaled wave packet is first propagated until $t=t_{\mathrm{final}}$, the end of the interaction of our model atom with the pulse. At time $t=t_{\mathrm{final}}$, the bound state contribution is subtracted from the total wave packet and scaling is switched on. At time $t=t_{\mathrm{final}}+250$ a.u., we clearly see on Fig. 4a that the scaled wave packet represented as a function of the position $\xi$ is confined compared to the unscaled one. In addition, the number of oscillations is already reduced. 
\begin{figure}[h]
\includegraphics[scale=0.23]{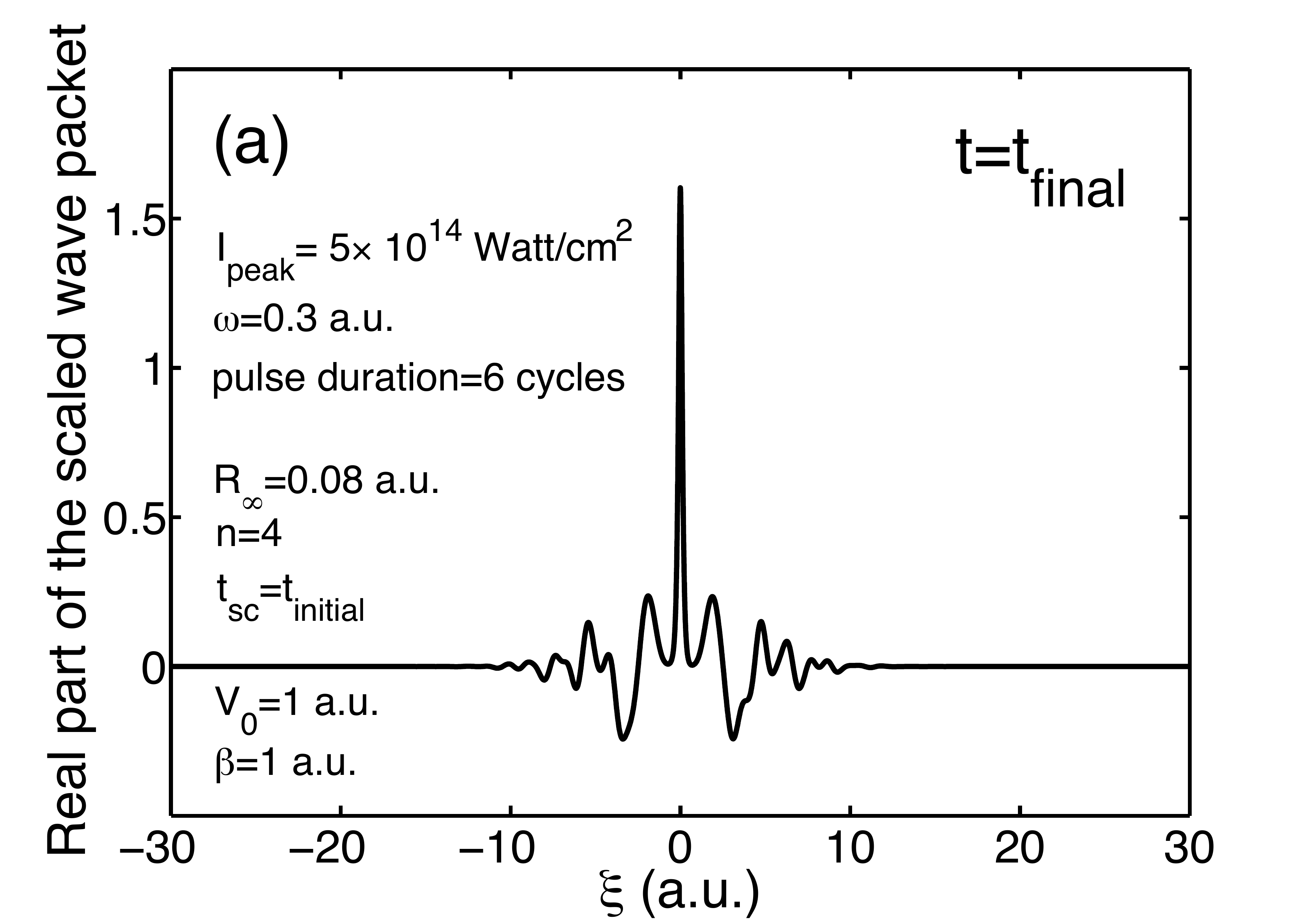} \includegraphics[scale=0.23]{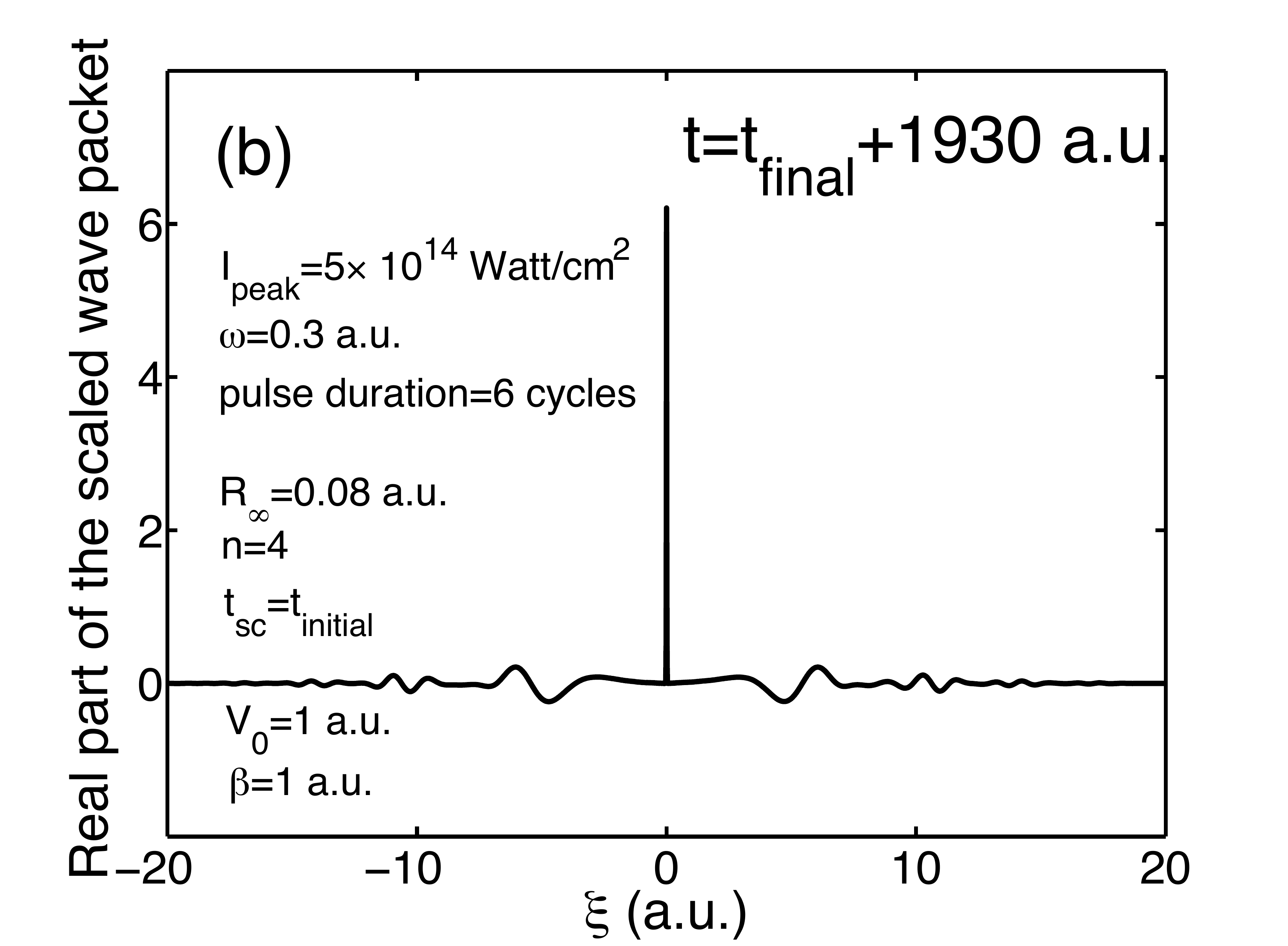}
\caption{Real part of the scaled wave packet resulting from the interaction of our one-dimensional model atom with a cosine square pulse of  $5\times 10^{14}$ Watt/cm$^2$ peak intensity and 0.3 a.u. photon energy. The total pulse duration is 6 optical cycles. The parameters of the Gaussian potential are the same as in Fig. 3. The time dependent scaling is switched on at the beginning of the interaction with the pulse. The asymptotic velocity $R_{\infty}$ is equal to 0.08 a.u. while the parameter $n$ of the scaling function (see Eq. 9) is equal to 4. The real part of the scaled wave packet is represented as a function of $\xi$ at (a) $t=t_{\mathrm{final}}$ and (b) $t=t_{\mathrm{final}}+1930$ a.u.. }
\end{figure}
As discussed above, the quadratic increase of the phase with the distance is canceled by the phase transformation (7) of the wave packet. In principle, the time scaling of the coordinates may start at any time.  In Fig.4b, we consider the same case as in Fig. 4a but the time scaling is now switched on right at the beginning of the interaction with the pulse. We clearly see that the confinement is slightly stronger and that the number of oscillations is significantly reduced. Indeed, beyond $\pm 120$ a.u., all the oscillations of weak amplitude present in Fig. 4a have disappeared. The comparison of Figs 4a and 4b also shows that the scaled wave packet has not yet reached a stationary state. In Fig. 5, we show the scaled wave packet resulting from the interaction of the same one-dimensional model atom as before with a pulse of $5\times 10^{14}$ Watt/cm$^2$ peak intensity and 0.3 a.u. photon energy. The total duration of the pulse is 6 optical cycles. In this particular case, we use B-splines to describe the scaled wave packet with two different exponential sequences of break points. The first grid with $\gamma=5$ (see Eq. 15) and 200 B-splines is 
used during the interaction of the model atom with the pulse. The second grid with $\gamma=5$ and 400 B-splines is used after the interaction with the pulse {\it i.e.} for $t>t_{\mathrm{final}}$. In Fig. 5, we show the real part of the scaled wave packet at the end of the pulse (Fig. 5a) and 1930 a.u. of time later when the scaled wave packet becomes quasi stationary. The fast oscillations have now been completely removed. It is important to stress that although the number of B-splines used is rather small, both grids allow us to describe accurately the shrinking of the ground state. In fact, for time $t\gg t_{\mathrm{final}}$ when the scaling function becomes linear with time, the scaled TDSE is identical to the original TDSE within a simple linear scaling of the spatial coordinates. This clearly shows that a multi-resolution approach in which the density of break points increases linearly with time around the origin makes sense. Note that in the present case, the accumulation of break points around the origin (right from the beginning of the propagation) increases significantly the stiffness of the system, a problem which should be avoided by means of multi-resolution techniques. The stiffness problem is discussed in the next section.

\section{Time propagation}

Solving accurately the TDSE usually requires the representation of  the solution on large or/and dense grids or  in large bases. In all cases, we deal with large systems of coupled first order differential equations which are well known to be stiff \cite{Huens97}. This means that the step size decreases as the dimension of the system increases. The origin of the stiffness is clear: by increasing the size or the density of the grid or the size of the basis, the Hamiltonian generates large positive eigenenergies, which are responsible for strong oscillations in the solution of the TDSE. It is thus the largest positive eigenvalue which controls the step size. In fact, the stiffness of the system may lead to the instability of the time propagation scheme  as well as to inaccurate high energy components of the solution \cite{Sein71}. Two approaches frequently used to overcome this problem are implicit schemes for solving the TDSE and the propagation of the TDSE in the atomic basis  and  possibly, within the interaction picture. The first method requires typically the solution of large systems of linear equations at each integration step. It is however important to stress that implicit schemes actually solve the stability problem but not necessarily the inaccuracy problem for the high energy components of the solution. In the second method, the time integration is achieved by means of explicit algorithms, which have been proved to be very stable for the solution of the TDSE in the atomic basis where the atomic Hamiltonian is diagonal. These algorithms only need matrix-vector products. However, the representation in the atomic basis requires the full diagonalization of the Hamiltonian before starting the integration. In any case, the computational cost increases dramatically with the size of the system. \\

It is therefore desirable to have an explicit algorithm suitable for the direct solution of stiff TDSE. Such a method does exist and was proposed more than thirty years ago by Fatunla \cite{Fatun78,Fatun80}. It has been successfully implemented for the description of single ionization of atoms by  strong oscillating fields \cite{Madro09,Madro10}. In this method that takes into account the intrinsic frequencies of the system, the wave function is expressed in terms of oscillating functions. This leads to a simple recursive formula for the time propagation with a controlled error. At each integration step, only matrix-vector products are therefore needed. In the two following subsections, we describe the most important features of Fatunla's algorithm and its implementation and show how its accuracy can be significantly improved within a predictor-corrector scheme.

\subsection{Fatunla's explicit scheme}
We start with the general matrix form of the TDSE using a spectral or a grid representation,
\bb
 \label{eq:tdsegen}
 {\rm i} \B\frac{\mathrm{d} \Bpsi}{\mathrm{d} t}=\H(t)\Bpsi,
\ee
with $\H(t)$ the matrix representation of the Hamiltonian, and $\B$ the overlap matrix, which is the identity in a grid representation or in an orthonormal basis.
Truncation of the basis or of the grid leads to a $m$-dimensional first order differential equation,
\bb
\label{bc:diffeq}
\dot{\Bpsi}=\f(t,\Bpsi),
\ee
where $\f(t,\Bpsi)=-{\rm i}\B^{-1}\H(t)\Bpsi$ is in general a complex $m$-dimensional function.\\

The stiffness  of equation (19) leads to a solution $\Bpsi(t)$ which is an oscillating function. In a given interval $[t_n,t_{n+1}],~t_{n+1}=t_n+h,$ with $h$ a small number, $\Bpsi(t)$ is approximated by the function
\bb
 \label{eq:ansatz}
 \F(t)=(\I-\mathrm{e}^{\Om_1\,t})\a-(\I-\mathrm{e}^{-\Om_2\,t})\b+\c,
\ee
with $\I$ the identity matrix, $\Om_i={\rm diag}(\omega_{1}^{(i)},\dots,\omega_{m}^{(i)})$, $i=1,2$, and $\a,\b,\c$ constant vectors. The complex numbers $\omega_{1}^{(i)},\dots,\omega_{m}^{(i)}$, $i=1,2$ are called {\em stiffness parameters}.
Assuming that $\F(t)$ coincides with $\Bpsi(t)$ at $t_n$ and $t_{n+1}$, that $\F'(t)$ coincides with $\f(t,\Bpsi)$ at $t_n$, and that $\F''(t)$ coincides with $\f'(t,\Bpsi)$ at $t_n$,  the solution  \hbox{$\Bpsi_{n+1}=\Bpsi(t_{n+1})$} at $t_{n+1}$ can be expressed recursively in terms of \hbox{$\Bpsi_{n}=\Bpsi(t_{n})$}, \hbox{$\f_n=\f(t_{n},\Bpsi_n)$} and \hbox{$\f_n^{(1)}=\mathrm{d}\f/\mathrm{d}t|_{t=t_{n}}$} according to
\bb
\label{eq:recursion}
 \Bpsi_{n+1}=\Bpsi_n+\R\f_n+\S\f_n^{(1)}.
\ee
$\R$ and $\S$ are diagonal matrices which can be written in terms of the stiffness parameters:
\bb
 \R=\Om_2\BPhi-\Om_1\BXi,\qquad \S=\BPhi+\BXi,
\ee
where $\BPhi$ and $\BXi$ are diagonal matrices of which the nonzero entries are 
\bb
\label{eq:phi}
 \Phi_i=\frac{\mathrm{e}^{\omega_i^{(1)}h}-1}{\omega_i^{(1)}(\omega_i^{(1)}+\omega_i^{(2)})}
\ee
and 
\bb
 \label{eq:psi}
 \Xi_i=\frac{e^{-\omega_i^{(2)}h}-1}{\omega_i^{(2)}(\omega_i^{(1)}+\omega_i^{(2)})}.
\ee
Notice that if a stiffness parameter, $\omega_i^{(1)}$  ($\omega_i^{(2)}$), vanishes the associated matrix element reads
\bb
 \Phi_i=\frac{h}{\omega_i^{(2)}} \quad\left(\Xi_i=-\frac{h}{\omega_i^{(1)}}\right).
\ee
The recursive relation (\ref{eq:recursion}) depends on the so far unknown stiffness parameters. These can be written in terms of the function $\f(t_n,\Bpsi_n)$ and its time derivatives $\f_n^{(k)}$, $k=0,1,2,3$, at $t_n$ \cite{Fatun80},
\begin{eqnarray}
\label{eq:stiffparam}
 \omega_i^{(1)}&=&\frac12\left[-D_i+\sqrt{D_i^2+4E_i}\right]\nonumber\\
 \omega_i^{(2)}&=&\omega_i^{(1)}+D_i,
\end{eqnarray}
where  $D_i$ and $E_i$, $i=1,\dots,m$ are given in terms of the respective components $f_{in}^{(k)}$ of $\f_n^{(k)}$, $k=0,1,2,3$ at $t=t_n$ by
\begin{eqnarray}
\label{eq:di}
 D_i&=&
\frac{f_{in}^{(0)}f_{in}^{(3)}-f_{in}^{(1)}f_{in}^{(2)}}{f_{in}^{(1)}f_{in}^{(1)}-f_{in}^{(0)}f_{in}^{(2)}},\quad i=1,\dots,m,\\
\label{eq:ei}
 E_i&=&
\frac{f_{in}^{(1)}f_{in}^{(3)}-f_{in}^{(2)}f_{in}^{(2)}}{f_{in}^{(1)}f_{in}^{(1)}-f_{in}^{(0)}f_{in}^{(2)}},\quad i=1,\dots,m,
\end{eqnarray}
provided that the denominator of the previous expressions is nonzero. It is important to note that in the present case, the  time-dependent scaling function (9) is considered as a parameter. As a result, the successive time derivatives of this function are not taken into account in the calculation of $D_i$ and $E_i$.\\

The $i$th component of the local truncation error at $t=t_{n+1}$, that is the difference between the exact solution at $t_{n+1}$ and the numerical solution, is given by (we ignore here the index $i$ of the components for the sake of clarity)
\cite{Fatun80}:
\begin{eqnarray}
 \label{eq:truncerror}
 T_{n+1}&=&\frac{h^5}{5!}\frac1{\omega_1+\omega_2}\left[ (\omega_1+\omega_2)f_n^{(4)}
+(\omega_2^4-\omega_1^4) f_n^{(1)}\right.\nonumber\\
&&\left.-(\omega_1^4\omega_2+\omega_1\omega_2^4)f_n^{(0)}\right]+{\cal O}(h^6)\nonumber\\
&=&\frac{h^5}{5!}\left[f_n^{(4)}+(\omega_2^3-\omega_2^2\omega_1+\omega_2\omega_1^2-\omega_1^3) f_n^{(1)}\right.\nonumber\\
&&\left.-\omega_1\omega_2(\omega_1^2-\omega_1\omega_2+\omega_2^2)f_n^{(0)}\right]+{\cal O}(h^6).
\end{eqnarray}

The implementation of the recursion (\ref{eq:recursion}) is now rather simple. It requires the calculation of the function $\f_n$ and its derivative $\f_n^{(1)}$ at each value of $t_n$. For the stiffness matrices $\Om_1$ and $\Om_2$, and thus also for the matrices $\R$ and $\S$, the derivatives $\f_n^{(2)}$ and $\f_n^{(3)}$ are also needed. $\Om_1$ and $\Om_2$ have to be calculated in principle at each integration step, since they characterize the local frequencies of the solution $\Bpsi(t)$. The truncation error (\ref{eq:truncerror}) can be used to control the size of the integration step, e.g., by imposing a boundary criterion for $|T_n|$. For this also, the derivative $\f_n^{(4)}$ must be provided.\\

\begin{figure}[h]
\begin{center}
\includegraphics[width=11cm,height=8cm]{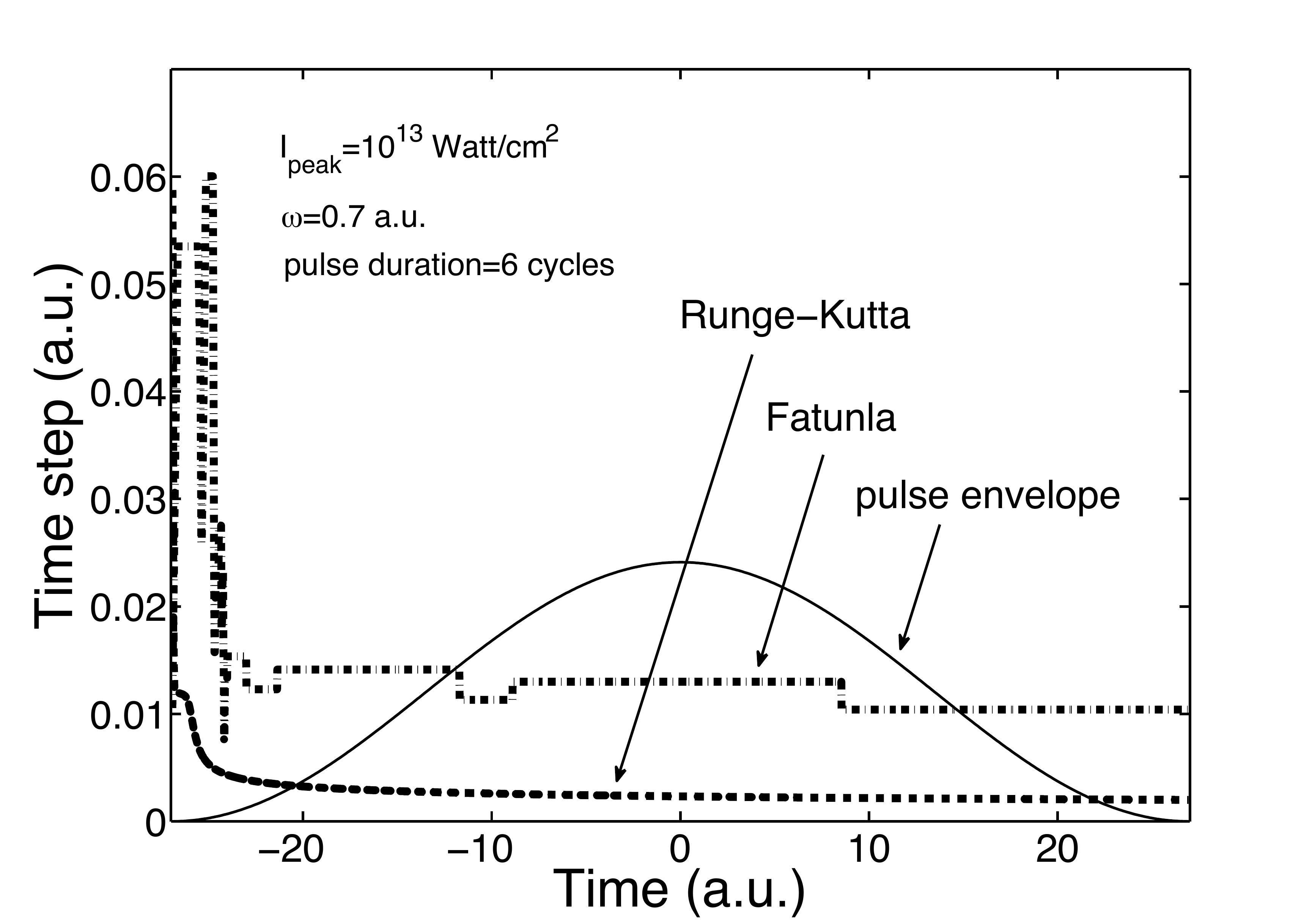} 
\caption{Comparison of the time steps used in Fatunla's method (dashed line) and a method based on a fifth order embedded Runge-Kutta formula (dotted line) to solve the TDSE for the same case as in Fig. 3. The time propagation is performed in a basis of 500 Hermite Sturmian functions of dilation parameter $\alpha=0.5$ a.u.. For the sake of completeness,  we also show the pulse envelope.}
\end{center}
\end{figure}
The stiffness parameters carry the intrinsic information of the natural oscillations of the system. Therefore, the time step is expected to be rather large compared with standard explicit methods, like Runge-Kutta. In order to illustrate this, we  solve without any time-dependent scaling of the coordinates, the TDSE (18) for our one-dimensional model atom ($V_0=1$ and $\beta=1$) interacting with a 6-cycles cosine square pulse of $10^{13}$ Watt/cm$^2$ peak intensity and 0.7 a.u. photon energy (same case as in Fig. 3). The time propagation is performed in the Hermite Sturmian basis by means of two approaches : Fatunla's method and a fifth-order embedded Runge-Kutta  formula \cite{Hairer87}. The Hermite Sturmian basis contains 500 functions with the dilation parameter $\alpha=0.5$ a.u.. In Fig. 6, we compare the time variations of the time step which is automatically adjusted in both methods. In the case of Fatunla's method, this is done according to the condition $10^{-16}\le T_n\le 10^{-12}$ for the truncation error. In the case of the embedded Runge-Kutta formula, the time step is adjusted by comparing the results obtained with the fifth order  formula and with a fourth order one which uses the same intermediate mesh points. We clearly see that the time step is bigger in the case of Fatunla's method when compared to the Runge-Kutta embedded formula. However, the relative error on the conservation of the norm of the wave packet is, in this particular case, two orders of magnitude better in the case of the Runge-Kutta embedded formula. In fact, by increasing the value of the dilation parameter $\alpha$ as well as the number of basis functions, the stiffness of the system of equations becomes more pronounced. In these conditions, Fatunla's method becomes much more efficient in terms of the magnitude of the time step and the relative error on the conservation of the norm is of the same order of the one obtained with the Runge-Kutta embedded formula.

\begin{figure}[h]
\begin{center}
\includegraphics[width=11cm,height=8cm]{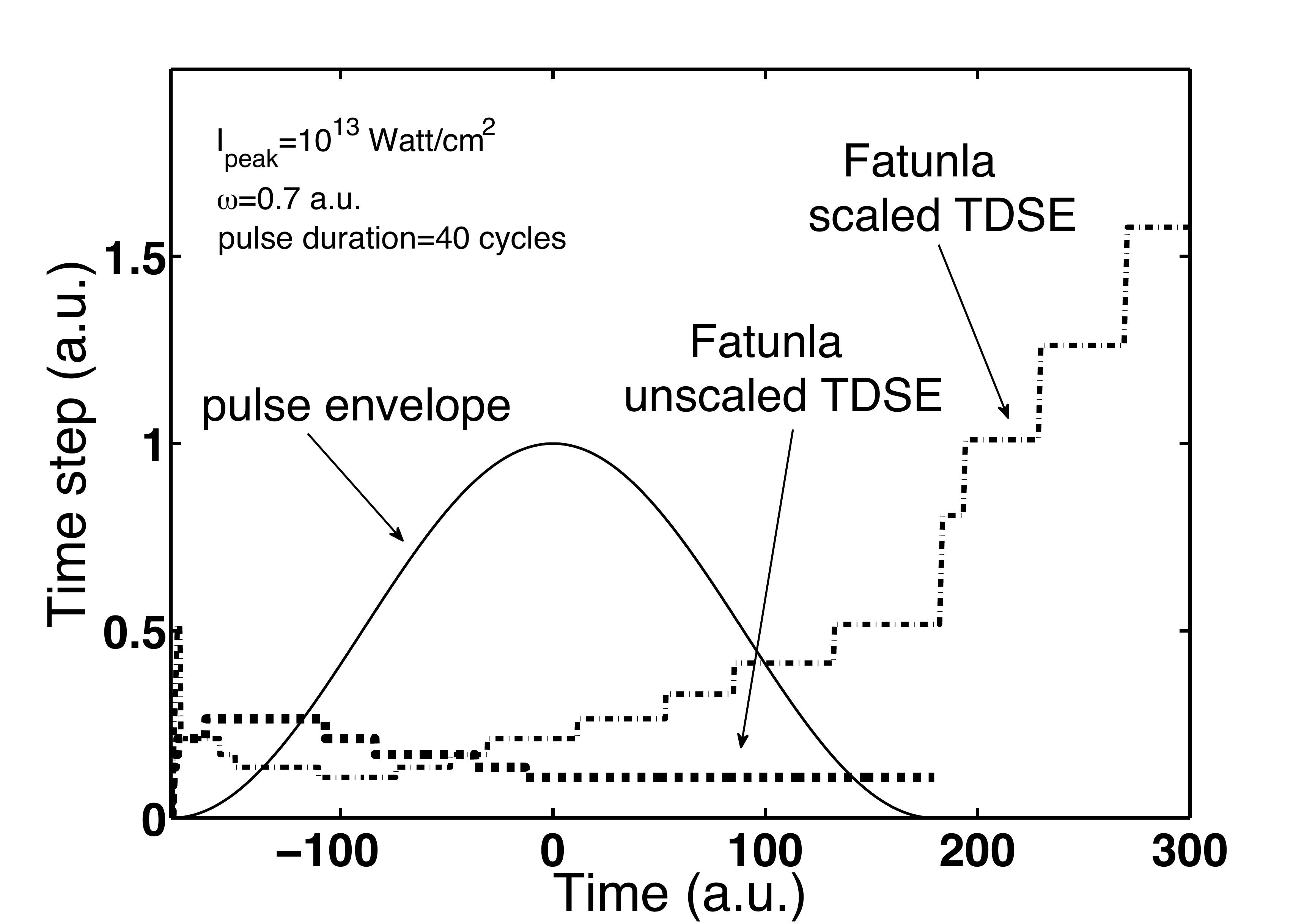} 
\caption{Comparison of the time steps used in Fatunla's method to solve the scaled TDSE (light dotted line) and the unscaled TDSE (thick dotted line) in the case of the interaction of our one-dimensional model atom ($V_0=1$ and $\beta=1$) with a cosine square pulse of  $10^{13}$ Watt/cm$^2$ peak intensity and 0.7 a.u. photon energy. The total pulse duration is 40 cycles. The scaled and unscaled TDSE are solved in a Hermite Sturmian basis. In the case of the scaled TDSE, 400 Hermite Sturmian functions are used. These functions have a dilation parameter $\alpha=0.05$ a.u..  The parameter $n$ of the scaling function is equal to 4 and the asymptotic velocity $R_{\infty}=0.01$ a.u.. For the unscaled TDSE, a basis of 2000  functions with a dilation parameter $\alpha=0.008$ a.u. is necessary. For the sake of completeness,  we also show the pulse envelope.}
\end{center}
\end{figure}
Fatunla's method has also been tested in much more demanding cases namely the interaction of atomic hydrogen with intense low frequency pulses. The corresponding TDSE has been solved in a Coulomb Sturmian basis. The results obtained with Fatunla's method are in very good agreement with those obtained with a fully implicit Radau method of order 7 (see reference \cite{Huens97,Madro10} for details). In fact, Fatunla's method turned out to be 10 times faster than the fully implicit method. However, the relative error on the conservation of the norm of the solution is hardly lower than $10^{-5}$ by contrast with the fully implicit method where the norm is perfectly conserved.\\

Let us now examine how Fatunla's method performs in the case of the scaled TDSE. We consider the interaction of our one-dimensional model atom ($V_0=1$ a.u. and $\beta=1$ a.u.) with a cosine square pulse of $10^{13}$ Watt/cm$^2$ peak intensity and 0.7 a.u. photon energy. The total pulse duration is equal to 40 optical cycles. In Fig. 7, we show how the time step used by Fatunla's method varies with time for the scaled and unscaled TDSE. We work in the Hermite Sturmian basis. In the case of the scaled TDSE, 400 basis functions with a dilation parameter $\alpha=0.05$ a.u. are sufficient. The parameter $n$ of the scaling function is equal to 4 and the asymptotic velocity $R_{\infty}=0.01$ a.u.. It is important to note that in this case, it is necessary to integrate over about 4000 a.u. of time to reach the region where the scaled wave packet becomes static. By contrast to the unscaled equation, we clearly see that the time step needed for the time propagation of the scaled wave packet increases significantly as soon as the interaction with the pulse has ceased. In fact, around $t=4000$ a.u., the time step has a value around 10 a.u.. This results from the subtraction of the scaled bound state at the end of the pulse and from the slow time evolution of the ionized wave packet in the scaled representation. It is important to stress here that the shrinking of the bound states will never stop whereas the ionized part of the wave packet becomes static at large times. It is therefore important to subtract the scaled bound states from the total wave packet. Note that the subtraction has to be performed when both  the interaction with the pulse and the harmonic potential have disappeared.\\

From the previous discussion, it turns out that Fatunla's method with adaptive step size is particularly well adapted to the solution of the scaled TDSE. In very stiff problems, the time step is much bigger than in the case of usual explicit Runge-Kutta methods. However, in all cases we have treated so far, the relative error on the conservation of the norm is of the order of $10^{-5}$ even when we force the time step to be very small. In many cases, this accuracy is sufficient but, there are always cases where a higher accuracy is needed. In the next subsection, we show that this accuracy problem is solved by using a predictor-corrector scheme in which Fatunla's method is the predictor.

\subsection{Predictor-corrector scheme}
Predictor-corrector (P-C) methods are pairs of an explicit and an implicit multistep method where the explicit formula is used to predict the next approximation and the implicit formula to correct it. The order of the implicit method is usually the same or higher than the order of the explicit method. In the present case, the predictor is Fatunla's method which is of order 5. The corrector is a fully implicit 4-stage Radau method of order 7 \cite{Butcher64}. The implementation of this implicit method within the P-C scheme follows Refs. \cite{Houwen91} and \cite{Houwen93}. It is based on diagonally implicit iterations that have two main advantages: it preserves the favourable stability characteristics of the fully implicit Radau method and it is suitable for use on parallel processors. In the following, we give a brief outline of the method.
Let $t_n$ and $t_{n+1}$, be two consecutive times at which we want to calculate the wave packet. The solution $\Bpsi_{n+1}\equiv\Bpsi(t_{n+1})$ of the TDSE is obtained from $\Bpsi_n$ through the following relation:
\bb
\Bpsi_{n+1}=\Bpsi_n+h\sum_{i=1}^4b_i\H(t_n+c_ih)\By_i,
\ee
where $h=t_{n+1}-t_n$ and $b_i$ and $c_i$ are coefficients that define Radau's method. $t_i=t_n+c_ih$ with $i$ varying from 1 to 4 are intermediate times in the interval $[t_n,t_{n+1}]$ with $t_4=t_{n+1}$ $(c_4=1)$. In the case of an orthonormal basis, $\H$ is the Hamiltonian matrix. When the basis is not orthonormal, $\H$ represents the Hamiltonian matrix multiplied on the left by the inverse of the overlap matrix. $\By_i$ is an estimation of $\Bpsi(t_i)$ which, within the fully implicit 4-stage Radau method, is obtained by solving the following system:
\bb
\left(\begin{array}{c}
           \By_1\\ \vdots \\ \By_4
        \end{array}\right)=\left(\begin{array}{c}
                                                    \Bpsi_n \\ \vdots \\ \Bpsi_n
                                                \end{array}\right)+h\left(\begin{array}{ccc}
                                                                                       a_{11}\H(t_1) & \cdots & a_{14}\H(t_4) \\
                                                                                       \vdots & \ddots & \vdots \\
                                                                                       a_{41}\H(t_1) & \cdots & a_{44}\H(t_4)
                                                                                   \end{array}\right)\left(\begin{array}{c}
                                                                                                                             \By_1\\ \vdots \\ \By_4
                                                                                                                         \end{array}\right).
\ee
where the coefficients $a_{ij}$ are given. They define like $b_i$ and $c_i$ the implicit 4-stage Radau formula. If $\H$ is a $m\times m$ matrix, the dimension of the above system is $4m$. Usually, when an implicit method is the corrector in a P-C scheme, the vector that contains the $\By_i$ in the right hand side of Eq. (31) is provided by the predictor and system (31) is solved iteratively. Those iterations are called explicit since they only require matrix-vector multiplications. However, for stiff problems, such an explicit iterative process does not always converge. Instead, we introduce the diagonal matrix $\Bd$=diag$(d_{11},d_{22},d_{33},d_{44})$ and rewrite system (31) as follows:
\begin{eqnarray}
\left(\begin{array}{c}
           \By_1^{(j)}\\ \vdots \\ \By_4^{(j)}
        \end{array}\right)&-&h\left(\begin{array}{ccc}
                                                           d_{11}\H(t_1) & \cdots & 0 \\
                                                            \vdots & \ddots & \vdots \\
                                                            0 & \cdots & d_{44}\H(t_4)
                                                        \end{array}\right)\left(\begin{array}{c}
                                                                                                  \By_1^{(j)}\\ \vdots \\ \By_4^{(j)}
                                                                                               \end{array}\right)\nonumber\\
&=&\left(\begin{array}{c}
           \Bpsi_n\\ \vdots \\ \Bpsi_n
        \end{array}\right)+h\left(\begin{array}{ccc}
                                                      (a_{11}-d_{11})\H(t_1) & \cdots & a_{14}\H(t_4) \\
                                                       \vdots & \ddots & \vdots \\
                                                       a_{41}\H(t_1) & \cdots & (a_{44}-d_{44})\H(t_4)
                                                    \end{array}\right)\left(\begin{array}{c}
                                                                                              \By_1^{(j-1)}\\ \vdots \\ \By_4^{(j-1)}
                                                                                           \end{array}\right),
\end{eqnarray}
where the superscript $j$ gives the order of the iteration. $j=0$ corresponds to the result obtained with Fatunla's method. Although systems (31) and (32) are perfectly equivalent since matrix $\Bd$ is subtracted on both sides of Eq. (31), we have now to solve four $m\times m$ systems of equations at each iteration. Those iterations are called implicit. Note that due to the diagonal structure of matrix $\Bd$, the four systems can be solved in parallel. Such an implicit iterative process converges rapidly requiring a few iterations. The choice of matrix $\Bd$ is in principle arbitrary. In the present case, it is calculated in order to preserve the stability properties of the fully implicit Radau method. See Ref. \cite{Houwen91} for more details. The present P-C method allows a perfect conservation of the norm. Furthermore, we checked that the four systems of equations can be easily solved by means of the biconjugate gradient algorithm which only requires matrix vector multiplications. Therefore, although the present P-C algorithm is of an implicit nature, it is easy to implement on parallel processors.\\

\section{Electron energy spectrum}
One of the main advantages of the TSC method is the fact that the electron energy spectrum may be expressed directly in term of the scaled wave packet at large times. Here, we show it explicitly in the case of our one-dimensional model and for atomic hydrogen.
\subsection{Analytical expression for the electron energy spectra}
\subsubsection{One-dimensional model}
The unscaled wave packet that is the solution of the TDSE (1) can always be written as follows:
\bb
\Psi(x,t)=\sum_na_n(t)\psi_n(x)\mathrm{e}^{-\mathrm{i}E_nt}+\int^{\infty}_{-\infty}c_k(t)\psi_k(x)\mathrm{e}^{-\mathrm{i}\frac{k^2}{2}t}\mathrm{d}k,
\ee
where $\psi_n(x)$ is the wave function associated to a bound state of our model atom with $E_n$ its energy. $a_n(t)$ represents the probability amplitude for the system to be in the corresponding bound state. $\psi_k(x)$ is the wave function associated to a continuum state of energy $E=k^2/2$ while $c_k(t)$ is the corresponding probability amplitude. $k$ is the wave vector: a positive value of $k$ corresponds to a wave propagating to the right while a negative value corresponds to a propagation to the left. The continuum wave functions satisfy the following orthogonality relation:
\bb
\int_{-\infty}^{\infty}\psi_k^*(x)\psi_{k'}(x)\mathrm{d}x=\delta(k-k').
\ee
In order to calculate the electron energy spectrum given by $|c_k(t\rightarrow\infty)|^2$, we let $x\rightarrow\pm\infty$ in Eq. (33). In this limit the first term of the right hand side of this equation goes to zero and,
\bb
\psi_k(x\rightarrow\infty)=\frac{1}{\sqrt{2\pi}}\mathrm{e}^{\mathrm{i}kx}.
\ee
Note that the limit $x\rightarrow -\infty$ leads to the same expression since $k$ is replaced by $-k$. As a result we have
\bb
\Psi(x\rightarrow\pm\infty,t)=\frac{1}{\sqrt{2\pi}}\int_{-\infty}^{\infty}c_k(t)\mathrm{e}^{\mathrm{i}t\left(\frac{k}{t}x-\frac{k^2}{2}\right)}\mathrm{d}k.
\ee
In order to calculate the spectrum, let us take the limit of the above integral  for $t\rightarrow\infty$. By using the stationary phase theorem, we obtain:
\bb
\Psi(x\rightarrow\pm\infty,t\rightarrow\infty)=c_k(t\rightarrow\infty)\frac{1}{\sqrt{\mathrm{i}t}}\mathrm{e}^{\mathrm{i}\frac{x^2}{2t}}.
\ee
where $k=x/t$ is the stationary phase point. From Eqs. (6) and (9), we write
\bb
k=\frac{x}{t}=R_{\infty}\left(1-\frac{t_{\mathrm{sc}}}{t}\right)\xi.
\ee
Using Eq. (7) that relates the scaled wave packet to the unscaled one, we finally get:
\bb
c_k(t\rightarrow\infty)=\sqrt{\mathrm{i}\left(\frac{1}{R_{\infty}}+\frac{t_{\mathrm{sc}}}{R(t)}\right)}\Phi\left(\frac{k}{R_{\infty}\left(1-\frac{t_{\mathrm{sc}}}{t}\right)},t\rightarrow\infty\right),
\ee
which establishes a direct link between the probability amplitude for the electron to be in the continuum with an energy $k^2/2$ and the scaled wave packet at large times. Below, we discuss results for the electron energy spectrum and compare them with those obtained by solving the unscaled TDSE. In this latter case, we calculate the energy spectrum by projecting the final wave packet on continuum pseudostates obtained by diagonalizing the atomic Hamiltonian. Note that we get the same result by projecting onto plane waves because the phase shift introduced by the short range Gaussian potential is negligible.

\subsubsection{Atomic hydrogen}
As in the previous case, the unscaled wave packet can be written as follows:
\bb
\Psi(\vec{r},t)=\sum_{n,l,m}a_{n,l}(t)\psi_{n,l}(r)Y_{l,m}(\theta,\phi)\mathrm{e}^{-\mathrm{i}E_nt}+\sum_{l,m}\int_0^{\infty}c_{lk}(t)R_l(kr)Y_{l,m}(\theta,\phi)\mathrm{e}^{-\mathrm{i}\frac{k^2}{2}t}k\mathrm{d}k,
\ee
where $\psi_{n,l}(r)$ is the radial wave function of an hydrogen bound state of energy $E_n$ and $a_{n,l}$ the corresponding probability amplitude. $Y_{l,m}(\theta,\phi)$ is a spherical harmonic that is function of the angular coordinates of vector $\vec{r}$. $R_l(kr)$ is the radial regular Coulomb function where $k$ is the magnitude of the electron momentum \cite{Mott65}. $c_{lk}(t)$ is the probability amplitude for the electron to be in the continuum. The asymptotic form of the radial regular Coulomb function is
\bb
R_l(kr)\underset{r\rightarrow\infty}{\rightarrow}N_l(k)\;\frac{1}{kr}\;\mathrm{sin}\left(kr-\frac{1}{2}l\pi+\frac{1}{k}\mathrm{ln}(2kr)+\sigma_l\right).
\ee
$N_l(k)$ is a normalization factor. In the present case, it is equal to $2\sqrt{k}$ because $R_l(kr)$ is normalized per unit energy. $\sigma_l$ is the Coulomb phase shift given by
\bb
\sigma_l=\mathrm{arg}\Gamma(l+1-\frac{\mathrm{i}}{k}),
\ee
where $\Gamma(x)$ is the gamma function. In the limit of large distances, the unscaled wave packet reduces to the following expression:
\bb
\Psi(\vec{r},t)\underset{r\rightarrow\infty}{\rightarrow}\sum_{l,m}\int_0^{\infty}\frac{\sqrt{k}}{\mathrm{i}r}\;c_{lk}(t)\;\mathrm{e}^{-\mathrm{i}\frac{k^2}{2}t}\;\left[\mathrm{e}^{\mathrm{i}\left(kr-\frac{l\pi}{2}+\sigma_l+\frac{1}{k}\mathrm{ln}(2kr)\right)}-\mathrm{e}^{-\mathrm{i}\left(kr-\frac{l\pi}{2}+\sigma_l+\frac{1}{k}\mathrm{ln}(2kr)\right)}\right]\;\mathrm{d}k.
\ee
Note that the sine function in the asymptotic expression (41) of the regular Coulomb function has been replaced by a sum of two complex exponentials that describe pure ingoing and outgoing spherical waves at large distances. In order to calculate the electron energy spectrum, we now examine the limit of the previous expression for $t\rightarrow\infty$. Before applying the stationary phase theorem, let us rewrite expression (43) as follows:
\bb
\Psi(\vec{r},t)\underset{r\rightarrow\infty}{\rightarrow}\sum_{l,m}\int_0^{\infty}\frac{\sqrt{k}}{\mathrm{i}r}\;c_{lk}(t)\;\mathrm{e}^{\mathrm{i}\left(\sigma_l+\frac{l\pi}{2}\right)}\;\mathrm{e}^{\mathrm{i}t\left(-\frac{k^2}{2}+\frac{kr}{t}+\frac{1}{kt}\mathrm{ln}(2kr)\right)}\;\mathrm{d}k.
\ee
In fact, it is easy to show that the ingoing spherical wave present in expression (43) does not contribute to the integral in the limit $t\rightarrow\infty$. In integral (44), the stationary phase point $k=k_0$ is the solution of the following equation:
\bb
-k+\frac{r}{t}-\frac{1-\mathrm{ln}(2kr)}{k^2t}=0.
\ee
To a  good approximation we can replace $k$ by $r/t$ in the third term of the left hand side of the above equation. As a result we obtain:
\bb
k_0=\frac{r}{t}-\frac{1-\mathrm{ln}(2\frac{r^2}{t})}{\frac{r^2}{t}}.
\ee
If $k_{\mathrm{e}}$ represents the velocity of the electron at large distances,  we can  show by means of classical mechanics, that for large times,
\bb
\frac{r}{t}\approx k_{\mathrm{e}}+\frac{1}{k_{\mathrm{e}}^2t}\mathrm{ln}(1+k_{\mathrm{e}}^3t).
\ee
Note that for large times, $k_0$, the stationary phase point coincides with $k_{\mathrm{e}}$. By using the stationary phase theorem and Eqs. (6), (7) and (9), we obtain:
\begin{eqnarray}
c_{lk_0}(t\rightarrow\infty)&=&\left[\mathrm{i}\left(\frac{1}{R_{\infty}}+\frac{t_{\mathrm{sc}}}{R(t)}\right)\right]^{-\frac{3}{2}}\left(\frac{r}{t}\right)\frac{\mathrm{e}^{-\mathrm{i}(\sigma_l-\pi/2)}}{\sqrt{k_0}}\exp{\left[\mathrm{i} \left(\frac{  \mathrm{ln}(2r^2/t)+2\frac{\mathrm{ln}(2r^2/t)}{r^2/t}}{\frac{r}{t}+\frac{\mathrm{ln}(2r^2/t)}{r^2/t}}\right)\right]}\nonumber\\
&\times& \Phi_l\left(\frac{r}{R_{\infty}(t-t_{\mathrm{sc}})},t\rightarrow\infty\right),
\end{eqnarray}
which for a given value of the angular momentum $l$, establishes a link between the probability amplitude for the electron to be in the continuum with an energy $E=k_{\mathrm{e}}^2/2\approx k_0^2/2$ and the corresponding $l$-component of the scaled wave packet and hence the electron 
energy spectrum. Our results for the energy spectrum are compared with those obtained by projecting the unscaled wave packet on Coulomb functions.\\

\subsection{Results and discussion}
As a proof-of-principle, we show in this section, that the present method provides very accurate results for the electron energy spectra 
at the expense of less computer resources than with the same propagation method without scaling. Here, we calculate two different electron energy spectra in rather demanding physical situations. First, we consider the ionization of our one-dimensional model atom with an intense low-frequency field. 
\begin{figure}[h]
\begin{center}
\includegraphics[width=14cm,height=10cm]{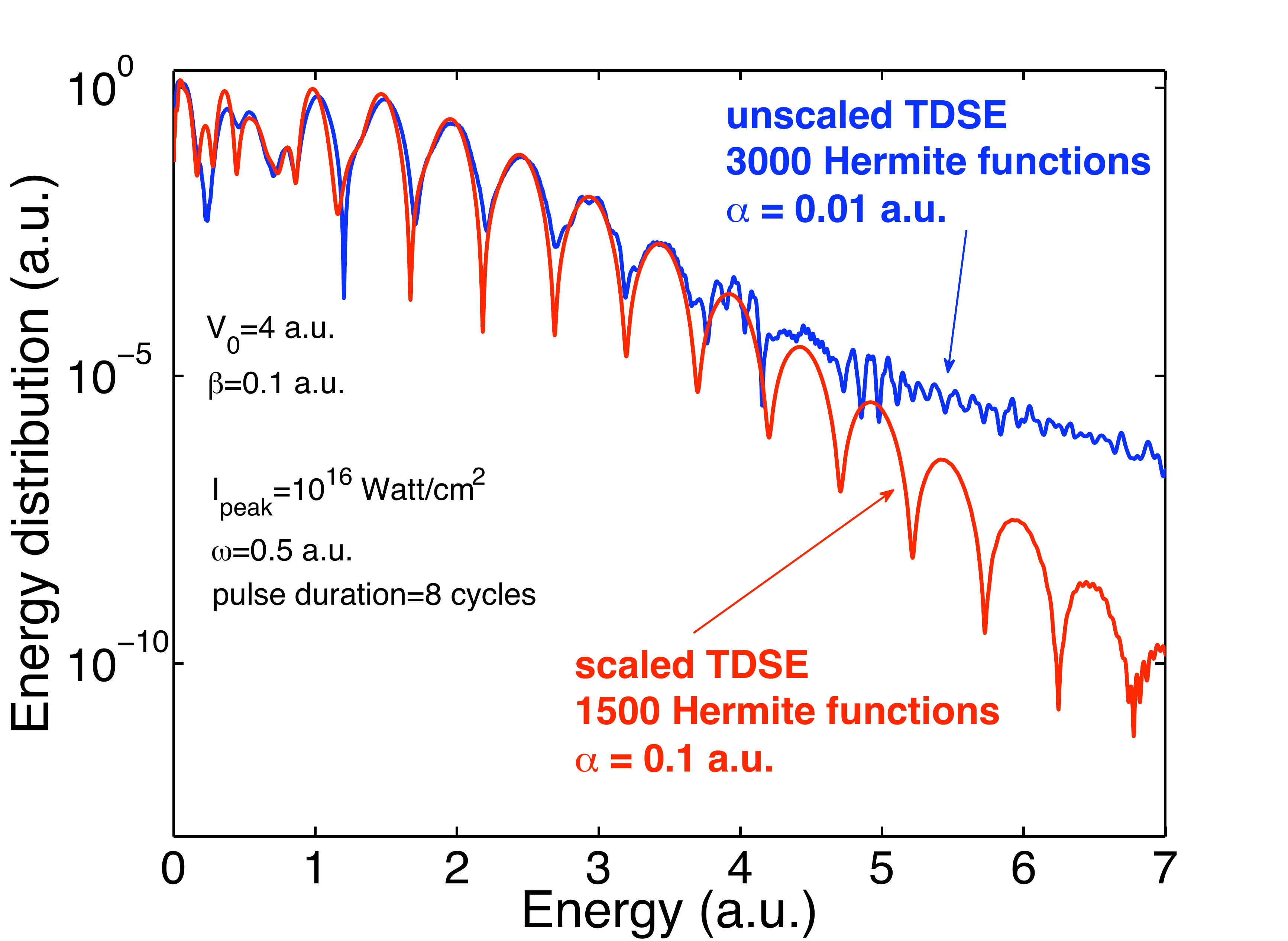} 
\caption{(Color online) Electron energy spectrum resulting from the interaction of our model atom with a strong low frequency pulse. The Gaussian potential parameters $V_0=4$ a.u. and $\beta=0.1$ a.u. are such that this potential supports seven bound states. The cosine square pulse of peak intensity, $10^{16}$ Watt/cm$^2$ and frequency $\omega=0.5$ a.u. has a total duration of 8 optical cycles. The full blue line is the result obtained by solving the unscaled TDSE with a basis of 3000 Hermite Sturmian functions and a dilation parameter $\alpha=0.01$ a.u.. The red full line is the result obtained by solving the scaled TDSE with a basis of 1500 Hermite Sturmian functions and a dilation parameter $\alpha=0.1$ a.u.. }
\end{center}
\end{figure}
The Gaussian potential parameters ($V_0=4$ a.u., and $\beta=0.1$ a.u.) are such that it supports seven bound states, the ground state energy being equal to -3.572 a.u.. The cosine square pulse has a total duration of 8 optical cycles with a peak intensity of $10^{16}$ Watt/cm$^2$ and a frequency of 0.5 a.u.. The time propagation of both the unscaled and scaled wave packets has been performed by means of the predictor-corrector method described above. The unscaled TDSE has been solved by expanding the wave packet in a basis of 3000 Hermite Sturmian functions with a dilation parameter $\alpha=0.01$ a.u.. The result (the full blue line) is shown in Fig. 8. We clearly see that the energy spectrum becomes noisy already around an electron energy of 3 a.u.. The reason is the following. Since fast electrons are emitted, the problem of the reflection of the unscaled wave packet by the numerical boundaries is a crucial issue. In order to overcome this problem, it is necessary to use a small value of the dilation parameter. 
In that case however, the density of positive energy states we obtain by diagonalizing the atomic Hamiltonian gets very small at high electron energy, thereby leading to the noisy behavior of the energy spectrum above 3 a.u.. The scaled TDSE has been solved by using a smaller basis of 1500 Hermite Sturmian functions with a dilation parameter of $\alpha=0.1$ a.u.. The correct spectrum (red full line) is obtained over more than 10 orders of magnitude after having time propagated the scaled wave packet over 5000 a.u. of time. The origin of the discrepancies observed at low electron energies is related to the choice of the asymptotic velocity. In the present case, $R_{\infty}=0.05$ a.u.. This value is in fact too high and leads to a strong confinement and thereby to a poor description of the shrinking of the bound states during the interaction with the pulse. When $R_{\infty}=0.03$ a.u., the discrepancies disappear and both curves are in good agreement at low electron energies. However, if the size of the basis stays equal to 1500, the result obtained by solving the scaled TDSE exhibits some unphysical oscillations at high energy. 
\begin{figure}[h]
\begin{center}
\includegraphics[width=11cm,height=8cm]{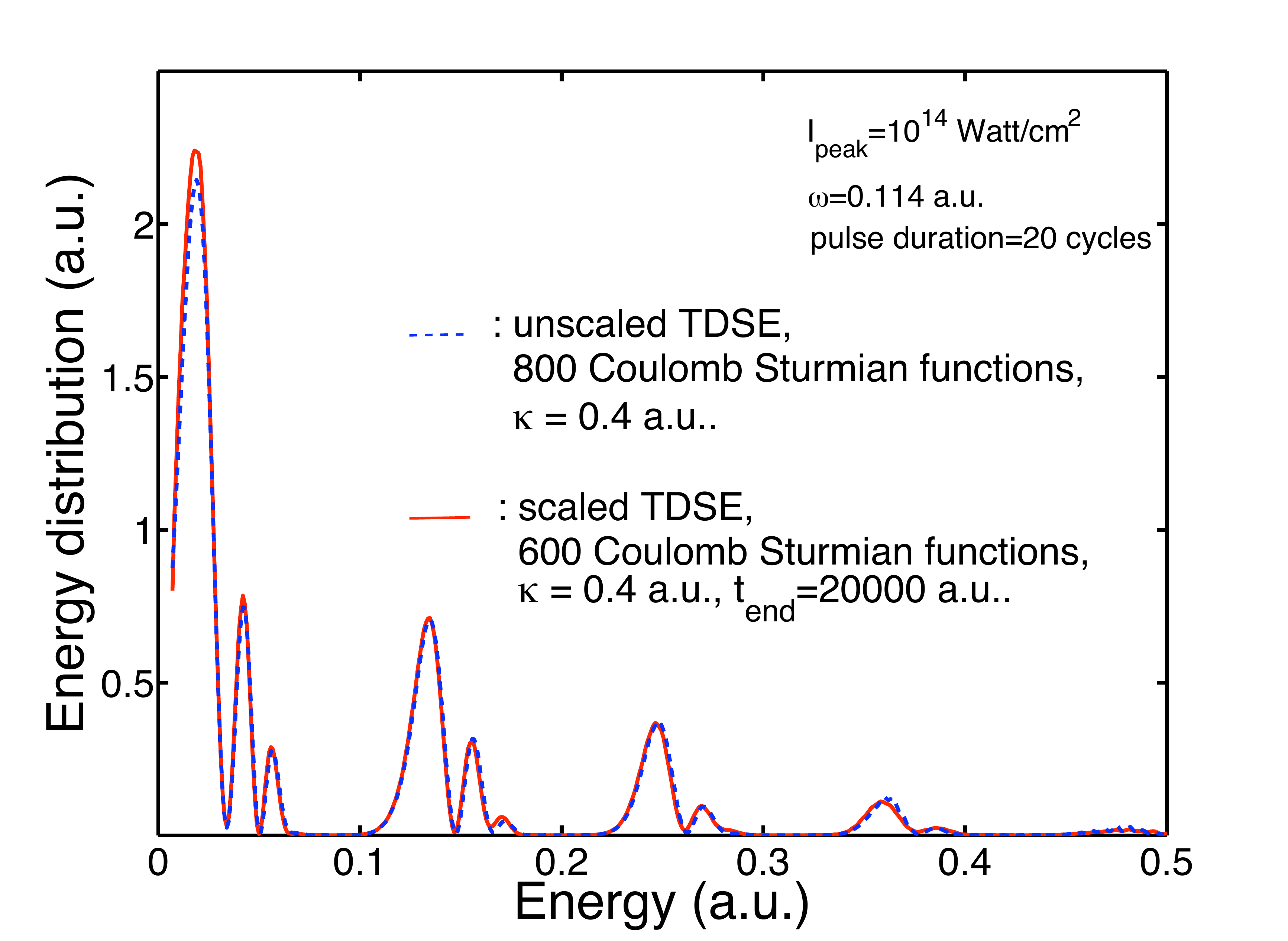} 
\caption{(Color online) Electron energy spectrum resulting from the interaction of atomic hydrogen with a strong low frequency laser field. The cosine square pulse of peak intensity $10^{14}$ Watt/cm$^2$ and frequency $\omega=0.114$ a.u. has a total duration of 20 optical cycles. The  blue dashed line is the result obtained by solving the unscaled TDSE with a basis of 800 Coulomb Sturmians per electron angular momentum and a dilation parameter $\kappa=0.4$ a.u.. The full red line is the result obtained by solving the scaled TDSE with a basis of 600 Coulomb Sturmian functions per electron angular momentum and the same dilation parameter. The scaled wave packet is propagated until $t_{\mathrm{end}}=20000$ a.u. of time where it is stationary.}
\end{center}
\end{figure}
In that case, the confinement is weaker requiring an increase of the size of the basis to cover a more extended region of space. An alternative could be to decrease the value of the dilation parameter. However, this dilation parameter fixes the spatial resolution in the whole space covered by the basis. It is therefore clear that a multi-resolution technique in which the resolution is increased  locally and gradually around the origin is more appropriate.\\

In Fig. 9, we consider the case of atomic hydrogen interacting with a 20 optical cycle cosine square pulse of peak intensity $10^{14}$ Watt/cm$^2$ and frequency 0.114 a.u.. The blue dashed line is the electron spectrum obtained by solving the unscaled TDSE with a basis of 800 Coulomb Sturmian functions with a dilation parameter $\kappa$ equal to 0.4 a.u.. This result is in perfect agreement with the one obtained by Grum-Grzhimailo {\it et al.} \cite{Grum10} (see Fig. 4 of that reference). The full red line is the electron spectrum obtained by solving the scaled TDSE with a basis of 600 Coulomb Sturmian functions with the same dilation parameter. Note that we have to propagate the scaled wave packet until $t_{\mathrm{end}}=20000$ a.u. {\it i.e.} during a long time after the end of the pulse. In fact, we have to wait until the wave packet becomes stationary before calculating the energy spectrum. In this context, it is therefore crucial to extract the scaled bound states. We clearly see that the results obtained by time scaling the coordinates are in perfect agreement with those obtained without scaling except at very small electron energies (first peak) where we observe a tiny difference which as before, is due to a slightly inaccurate description of the shrinking of the scaled bound states. In the present case, the scaling is switched on right at the beginning of the interaction and we use a small value of the asymptotic velocity, $R_{\infty}=0.001$ a.u., because of the long duration of the pulse. \\

\section{Conclusions and perspectives}
 In this contribution, we develop an {\it ab initio} approach to solve numerically the time-dependent Schr\"odinger equation that governs the ionization dynamics of atoms and molecules interacting with pulsed radiation fields. The approach is based on the combination of the time scaled coordinate method with an efficient time propagator. The key points of the time scaled coordinate method is a time-dependent scaling of the electron radial coordinate  together with a phase transformation of the total wave packet of the system. This method presents the following advantages: (i) the fast oscillations resulting from the rapidly growing phase gradients are removed from the total wave packet thanks to the phase transformation, (ii) the scaled wave packet stays spatially confined while reaching a stationary state a sufficiently long time after the interaction with the pulse and (iii) the electron energy distribution is proportional to the modulus square of the scaled wave packet once it becomes stationary. This method has however an important drawback: it introduces different length scales in the problem. In particular, it leads to a shrinking of the scaled bound states. In principle, such an effect can be described properly by using a denser grid or a much bigger basis of $\mathcal{L}^2$ functions. However, this inevitably increases the stiffness of the system of first order differential equations to solve for the time propagation of the scaled wave packet. Here, we show one efficient way of treating these problems. First, it is important to subtract the scaled bound states from the total wave packet after the end of the pulse, once the harmonic potential that confines the wave packet has disappeared. Second, we introduce a new high order time propagator of predictor-corrector type that so far revealed to be very efficient in handling the stiffness problem. The predictor is the fifth order explicit method of Fatunla and the corrector, a fully implicit Radau method of order seven. Despite the implicit character of the corrector, we show that all the calculations reduce to simple matrix-vector products that allow a high level parallelization of the computer codes. Finally, our calculations suggest that an elegant way to further improve the efficiency of our method is the use of multi-resolution techniques.\\
 
 At this stage, the method has been tested in the case of the interaction of a pulsed radiation field with a one-dimensional model atom described by a Gaussian potential and with atomic hydrogen. Electron energy spectra have been calculated in rather demanding physical situations. In all cases, the new approach give very accurate results, particularly for high photolectron energies, at the expense of less computer resources when compared to the usual grid or spectral methods without scaling.

\section*{Acknowledgements}

The authors enjoyed very interesting discussions with Laurence Malegat and Yuri Popov.
 J.E., J.M. and P.O'M thank the Universit\'e Catholique de Louvain (UCL) for financially
supporting a few stays at the Institute of Condensed Mater and Nanosciences (IMCN) of the UCL. J.E. greatfully acknowledges financial support by Deutsche Forschungsgemeinschaft under the contracts FR 591/16-1 and the Emmy-Noether group KR 2889/2. J.M. thanks the Deutsche Forschungsgemeinschaf for financial support under the contracts FR 591/16-1 and MA 3305/2-2.
B.P. thanks Royal Holloway College, University of London for hospitality and financial support.
A.L.F. gratefully acknowledges the financial support of the IISN (Institut Interuniversitaire des Sciences NuclŽaires) through the contract no 4.4.503.02.F,  ``Atoms, ions and radiation. Experimental and theoretical study of fundamental mechanisms governing laser-atom interactions and of radiative and collisional processes of astrophysical and thermonuclear relevance" . 
The authors thank UCL for providing them with access to the supercomputer of the CISM
(Calcul Intensif et Stockage de Masse) which is supported by the FNRS (Fonds National de la 
Recherche Scientifique) through the FRFC  (Fonds de la recherche fondamentale collective) 
project no 2.4556.99,  ``Si\-mulations Num\'eriques et traitement des donn\'ees".

\end{document}